\documentclass[%
 reprint,
superscriptaddress,
 longbibliography,
 amsmath,amssymb,
 aps,
prx,
floatfix,
]{revtex4-2}
\usepackage[colorlinks=true,
            linkcolor=blue,
            citecolor=blue,
            urlcolor=blue,
            filecolor=blue,      
            pdfborder={0 0 0}    
           ]{hyperref}
\usepackage{float} 
\usepackage{graphicx}
\usepackage{dcolumn}
\usepackage{esvect}
\usepackage{braket}
\usepackage{placeins}
\usepackage{physics}
\newcommand{\placeonpage}[2]{%
  \clearpage                       
  \begingroup
  \loop\ifnum\value{page}<#1
    \null\newpage                  
  \repeat
  #2\par
  \endgroup
  \FloatBarrier                    
}

\usepackage{bm}

\begin{document}

\preprint{APS/123-QED}

\title{Understanding Surface-Induced Decoherence of NV Centers in Diamond}
\author{Jonah Nagura}
    \affiliation{Pritzker School of Molecular Engineering, University of Chicago, Chicago, IL 60637, USA}
\author{Mykyta Onizhuk}
   \affiliation{Pritzker School of Molecular Engineering, University of Chicago, Chicago, IL 60637, USA}
\author{Giulia Galli}%
   \email{gagalli@uchicago.edu.}
   \affiliation{Pritzker School of Molecular Engineering, University of Chicago, Chicago, IL 60637, USA}
   \affiliation{Department of Chemistry, University of Chicago, Chicago, IL 60637, USA}
   \affiliation{Materials Science Division and Center for Molecular Engineering, Argonne National Laboratory, Lemont, IL 60439, USA}

\date{\today}

\begin{abstract}

Nitrogen vacancy centers (NV) in proximity of diamond surfaces are promising nanoscale quantum sensors. However, their coherence properties are negatively affected by magnetic and electric surface noise, whose origin and detailed impact have remained elusive. Using atomistic models of diamond surfaces derived with density functional theory, together with decoherence time calculations with cluster correlation expansion methods,  we quantify the effects of surface crystallographic orientation and functionalization, and of the density of unpaired electrons on the NV Hahn-echo time (T$_2$).  We determine a crossover depth at which $T_2$ ceases to be limited by surface-nuclear spins and recovers the bulk-limited value. We find that for static surface-electron baths, the ratio between the NV depth and the separation between surface electron spins determines a transition from fast-fluctuating to quasi-static noise, leading to a dependence of $T_2$ on orientation for specific surfaces. We also find that the modulation of $T_2$ by spin-phonon relaxations lead to motional‐narrowing at sub-microseconds relaxation times. Importantly, our calculations show that it is only when accounting for the surface spin in-sequence hopping that measured $T_2$ values as a function of depth can be reproduced, thus highlighting the  importance of hopping-mediated models to describe the surface spin noise affecting NV sensors. Overall, our work provides clear guidelines for engineering diamond surfaces to achieve enhanced NV coherence for quantum sensing and information processing applications.
\end{abstract}

\maketitle


\section{\label{sec:level1}Introduction}
Precise control of the coherence properties of quantum systems has enabled the development of novel  measurement techniques \cite{degen2017, taylor2008high}. For example, quantum sensing with ultra-high sensitivity \cite{balasubramanian2008nanoscale, degen2008scanning,   grinolds2014subnanometre} has been achieved utilizing spin defects in proximity to the surfaces of semiconductors or insulators. The negatively charged nitrogen‑vacancy (NV) center in diamond has emerged as an attractive spin defect for sensing applications, offering room‑temperature optical initialization, coherent spin control, and spin‑state‑dependent fluorescence readout \cite{doherty2013, schirhagl2014nitrogen}. However, shallow NV centers (those within $\sim$ 20 nm of a diamond surface) exhibit charge instability, reduced fluorescence, and increased spin dephasing \cite{ofori2012spin, rosskopf2014investigation, hauf2011chemical, Ohashi,  bluvstein2019identifying, yuan2020charge}. Noise spectroscopy experiments have shown that the decoherence arises from a complex interplay of surface-induced noise sources, including electric‑field fluctuations due to dynamic surface charges, magnetic noise from fluctuating surface spins, and surface phonons \cite{rosskopf2014investigation, romach2015spectroscopy, kim2015decoherence, jamonneau2016competition, myers2017double, candido2024interplay}. To mitigate noise, prior work has targeted electric-field fluctuations with dielectric coatings \cite{kim2015decoherence, chrostoski2018electric, lin2022diamond, candido2024theory} or atomic force microscope tips to manipulate surface charges\cite{zheng2022coherence} and magnetic fluctuations with surface treatments \cite{myers2014probing, chrostoski2021magnetic, kaviani2014proper, sangtawesin2019origins}, interface engineering \cite{lo2025enhancement, yu2025engineering}, and advanced quantum control protocols \cite{sushkov2014magnetic, bluvstein2019extending}.

Despite many experimental efforts, precise control of the chemical environment of diamond surfaces have not yet been achieved \cite{janitz2022diamond, chou2023ab}, and most noise spectroscopy investigations have not conclusively identified the species responsible for decoherence. Surface-sensitive X-ray spectroscopies and scanning tunneling microscopy of diamond surfaces have revealed unoccupied defect states (electron traps) and surface terminations bearing multiple nuclear-spin species, indicating a heterogeneous nuclear-spin landscape and a complex electron-spin environment \cite{stacey2019evidence, sangtawesin2019origins, sung2024identification}.
Double electron-electron resonance (DEER) measurements have detected paramagnetic centers with spin $S=1/2$ and a g-factor close to that of free electrons, consistent with the presence of surface defects with unpaired electrons \cite{grinolds2014subnanometre}. Recent experiments further suggest that some surface spins are mobile under green-laser illumination and exhibit finite relaxation times \cite{dwyer2022probing, de2018density}. These findings underscore the need to investigate the specific sources of noise on diamond surfaces, including the dynamical behavior of electronic spins, and their impact on NV coherence. A clearer understanding of the spin dynamics occurring at diamond surfaces is crucial to refine noise mitigation strategies and to define the fundamental limits of the sensitivity and resolution of diamond-based nanosensors.

In this work, we present a comprehensive study of surface‑induced decoherence of NV centers in diamond, based on a series of first principles calculations. We build and optimize atomistic models of diamond surfaces with varying crystallographic orientations and chemical terminations. Using density functional theory (DFT), we then compute the spin-coupling tensors of near surface NV centers. To simulate the center's decoherence properties, we develop an extended quantum bath model \cite{witzel2005quantum, yao2006theory} that incorporates both relaxation and dynamic hopping of surface spins. Our spin dynamics calculations are carried out  within the cluster-correlation expansion (CCE) framework \cite{witzel2006quantum, yang2008quantum,  onizhuk2024understanding}, applied to both closed and dissipative quantum systems.  Importantly, our approach enables the disentanglement of individual noise contributions, revealing the microscopic origins of varied surface-induced spin noise processes. Based on our results, we propose targeted surface-engineering and quantum control strategies to suppress dominant decoherence channels and we define the operational regimes in which quantum coherence of near surface NV centers can be optimized.

The paper is organized as follows. In Section II, we present our theoretical framework and introduce the NV decoherence model that incorporates both dissipation and hopping mechanisms. Section III presents our results, including a comparison of our predictions with available experimental data. Section IV concludes the paper.

\section{THEORETICAL AND COMPUTATIONAL FRAMEWORK}
\subsection{Spin Hamiltonian}
The spin Hamiltonian for a near‑surface NV center coupled to nuclear and electron  spin bath can be written as:

\begin{equation}\label{coh}
    \mathcal{\hat H} = \mathcal{\hat H}_{NV} + \mathcal{\hat H}_{NV-\mathcal{B}} + \mathcal{\hat H}_{\mathcal{B}}.
\end{equation}

Here, the NV center spin Hamiltonian $\mathcal{H}_{NV}$ includes the zero‑field splitting (axial D, and transverse E) terms, Zeeman coupling, and the Stark shifts due to external electric fields $(\mathcal{E})$ at the surface.

\begin{align}
\mathcal{ \hat H}_{NV} =\; & \gamma_{NV}\,{B}_z\cdot{S}_z 
+ D\Bigl(S_{z}^{2} - \frac{2}{3}\mathbb{I}\Bigr)
+ E\Bigl(S_{x}^{2} - S_{y}^{2}\Bigr) \nonumber\\[1mm]
& + d_{\parallel}\,\mathcal{E}_{z}\Bigl(S_{z}^{2} - \frac{2}{3}\mathbb{I}\Bigr) \nonumber\\[1mm]
& + d_{\perp}\,\mathcal{E}_{x}\Bigl(S_{x}^{2} - S_{y}^{2}\Bigr)
+ d_{\perp}\,\mathcal{E}_{y}\Bigl(S_{x}S_{y} + S_{y}S_{x}\Bigr),
\end{align}

where $d_{\parallel}$ and $d_{\perp}$ are the parallel and perpendicular electric-field susceptibilities of the NV center. 
The NV-bath coupling comprises hyperfine tensors ${\mathbf{A_i}}$ to nuclear spins and dipolar tensors  ${\mathbf{D_j}}$ to electron spins.

\begin{equation}
   \mathcal{\hat H}_{NV-\mathcal{B}}= \sum_{i \in\mathcal{N}_{nuc} }{\mathbf{S\; A_i\; I_{i}}} + \sum_{j \in \mathcal{N}_{e} }{\mathbf{S\;  D_j\;  S_j}},
\end{equation}

The bath Hamiltonian $\mathcal{ \hat H}_{\mathcal{B}}$ includes the Zeeman interaction of nuclear and electron spins, the nuclear quadrupole interaction ${\mathbf{Q}}$ for nuclear spins $(S \geq 1)$, and  inter‐bath‐spin interactions, inclusive of electron‑nuclear hyperfine couplings ${\mathbf{D_{ij}}}$ and dipolar couplings among the nuclear spins ${\mathbf{J_{ij}}}$ and electron spins ${\mathbf{P_{ij}}}$:

\begin{equation} \label{spin_ham}
\begin{split}
\mathcal{ \hat H}_{\mathcal{B}}= {} & \sum_{i \in \mathcal{N}_{nuc} } \gamma_{n,i}\,{B}_z{I}_z + \sum_{i \in \mathcal{N}_{nuc} } \mathbf{I_i\; Q\;I_i} + \sum_{\mathcal{\substack{i,j\in \mathcal{N}_{nuc}}}}{\mathbf{I_i\;  J_{ij}\; I_j}} \\ & + \sum_{i \in \mathcal{N}_{e}} \gamma_{e}\,{B}_z{S}_z  + \sum_{\mathcal{\substack{i,j\in \mathcal{N}_{e}}}}{\mathbf{S_i\;  D_{ij}\; S_j}} + \sum_{\substack{i \in \mathcal{N}_e \\ j \in \mathcal{N}_{nuc} }} \mathbf{S_i\;P_{ij} \;I_j}
\end{split}
\end{equation}

where, $\mathbf{S}=(S_x, S_y, S_z)$  and $\mathbf{I}=(I_x, I_y, I_z)$ are the electron and nuclear spin operators, respectively (See Appendix A for more details). 

\subsection{Dissipative Spin Baths}
To account for any local dissipation of the NV center and the spin-bath from a  Markovian environment, we compute the evolution of the total density matrix operator \(\hat{\rho}\) with a Lindblad master equation:
\begin{equation}\label{eq:lindblad}
\begin{split}
    \frac{d}{dt}\hat{\rho} =\; & -\frac{i}{\hbar}\Bigl[\mathcal{\hat{H}},\,\hat{\rho}\Bigr] + \Gamma_{NV}\,\mathcal{D}[\hat{L}](\hat{\rho}) \\
    & \quad + \sum_{i\in \mathcal{N}} \Gamma_i\,\mathcal{D}[\hat{L}_i](\hat{\rho})
    + \sum_{\substack{i,j\in \mathcal{N} \\ i\neq j}} \Gamma_{ij}\,\mathcal{D}[\hat{L}_{ij}](\hat{\rho})\,.
\end{split}
\end{equation}
where, \(\mathcal{\hat{H}}\) is given by Eq.~(\ref{coh}). \(\Gamma_{NV}\) is the NV dissipation rate; \(\Gamma_i\) and  $\Gamma_{ij}$ corresponds to the single-spin and pairwise incoherent bath processes, respectively. The super-operator $\mathcal{D}[\hat{L}]$ denotes the standard Lindblad dissipator:
\begin{equation}
\mathcal{D}[\hat{L}](\hat{\rho}) = \hat{L}\,\hat{\rho}\,\hat{L}^\dagger - \frac{1}{2}\Bigl\{\hat{L}^\dagger\hat{L},\,\hat{\rho}\Bigr\}\,.
\end{equation}

The jump operators \(\hat{L}\), \(\hat{L}_i\), and \(\hat{L}_{ij}\) account for the various incoherent processes (spin flips or dephasing) present at the surface. Details for the Lindblad jump operators are provided in Appendix B.

\subsection{Cluster‑Correlation Expansion}

The coherence of an NV center in a spin bath is computed via the cluster‑correlation expansion (CCE) method \cite{witzel2006quantum, yang2008quantum, onizhuk2024understanding}. We define the coherence function \(\mathcal{L}(t)\) as the normalized off‑diagonal element of the NV’s density matrix, which is factorized into irreducible contributions from various spin-bath clusters \(\mathcal{C}\):
\begin{equation}
\mathcal{L}(t)
=\prod_{\mathcal{C}}\tilde L_{\mathcal{C}}(t)
=\prod_i\tilde L_{\{i\}}(t)\,\prod_{i<j}\tilde L_{\{i,j\}}(t)\,\cdots.
\end{equation}

In conventional CCE, for each cluster \(\mathcal{C}\) with bath density \(\rho_{\mathcal{C}}\), one computes
\begin{equation}
L_{\mathcal{C}}(t)
=\mathrm{Tr}\bigl[U_{\mathcal{C}}^{(0)}(t)\,\rho_{\mathcal{C}}\,U_{\mathcal{C}}^{(1)\dagger}(t)\bigr],
\end{equation}
where \(U_{\mathcal{C}}^{(\alpha)}(t)\) propagates the bath conditioned on the NV state \(\ket{\alpha}\) (\(\alpha=0,1\)).

The generalized CCE (gCCE) \cite{onizhuk2021probing} instead retains the full NV Hilbert space in the cluster:
\begin{equation}
L_{\mathcal{C}}(t)
=\bra{0}\,U_{\mathcal{C}}(t)\,\rho_{\mathcal{C+S}}\,U_{\mathcal{C}}^\dagger(t)\,\ket{1},
\end{equation}
which captures non‑secular single and double‑quantum transitions and spin mixing from strain or electric fields. 

Coupling the Lindblad dynamics of (Eq.~\ref{eq:lindblad}) to CCE yields ME-CCE and ME-gCCE \cite{onizhuk2024understanding}, covering regimes from closed to dissipative baths. Together, these methods provide a flexible and accurate framework for simulating NV coherence dynamics across a broad range of experimental conditions.

\subsection{Atomistic Diamond Surface Models}
The NV coupling to magnetic signals depends strongly on the NV axis relative to the diamond surface \cite{bruckmaier2021geometry, janitz2022diamond}. Owing to their 
\(C_{3v}\) symmetry, the NV axis may be oriented along any of four equivalent \(\langle 111 \rangle\) directions. On conventionally grown (100) diamond these orientations occur with equal probability, but preferential alignment is achievable by controlling the surface orientation during growth \cite{pham2012enhanced, edmonds2012production, michl2014perfect, lesik2014perfect}. To isolate surface-orientation effects, we constructed atomistic models of (100), (110), (111), and (113) surfaces \cite{Widmann, LESIK201547, SHEN202217}. \par
\begin{figure*}[ht]
    \centering
    \includegraphics[width=\textwidth]{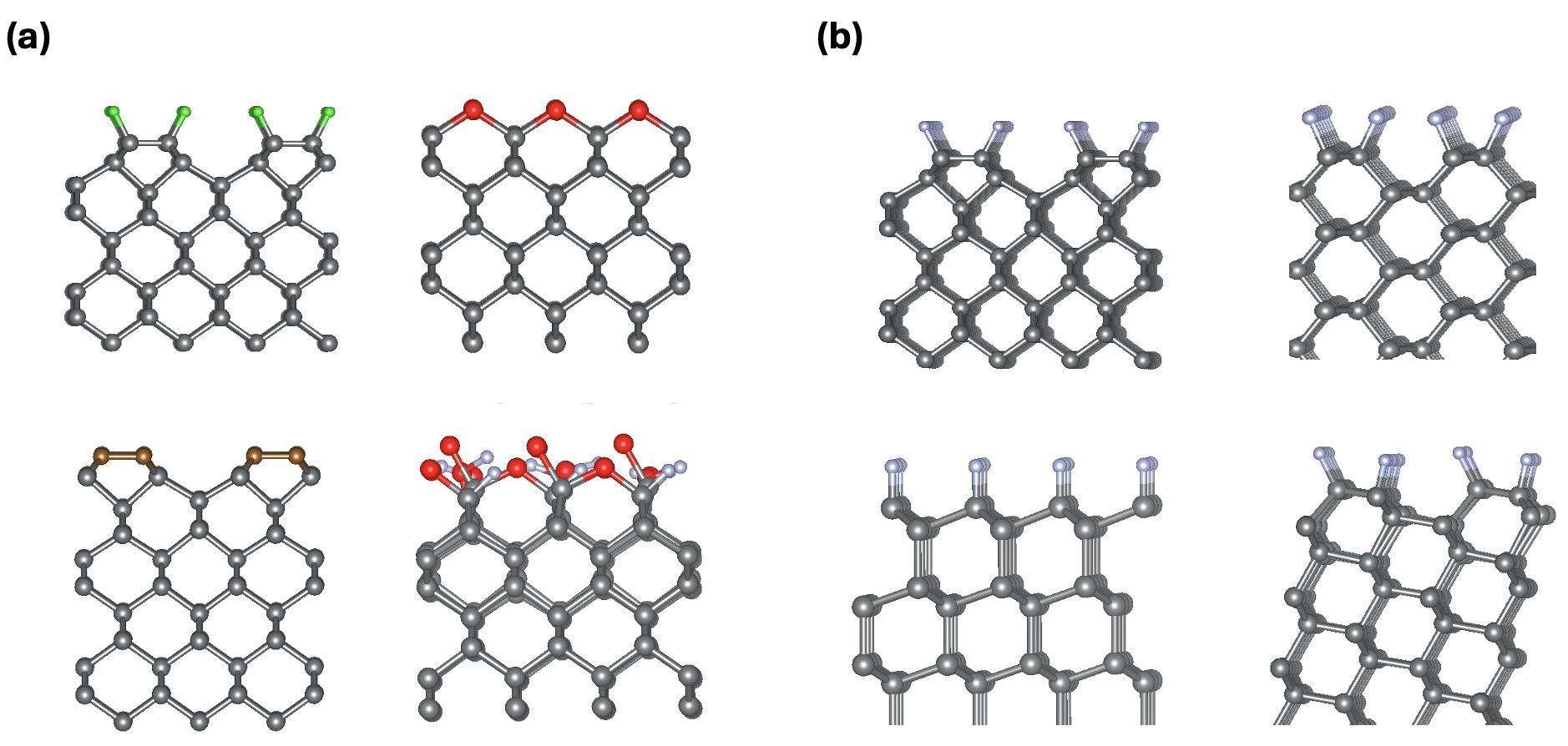}
    \caption{Atomistic models of diamond surfaces. \textbf{(a)} Representative reconstructions of the (100) surface with different terminating species; Fluorine (green sphere), Oxygen (red sphere), and Nitrogen (brown sphere)  were used to terminate a 2×1 reconstructed surface, while mixed hydroxyl/oxygen/hydrogen (OH/O/H) was used to terminate a 1×1 unreconstructed surface
\textbf{(b)} Hydrogen‐terminated surfaces for four distinct crystallographic orientations: (100), (110), (111), and (113).}
\label{fig:FIG1}
\end{figure*}

Our models are based on experimentally observed and thermodynamically stable \((2 \times 1)\) reconstructions, with terminations (O, H, F, and N) mimicking functionalizations used in experiments \cite{surf1, sangtawesin2019origins, Kawai, kaviani2014proper, SHEN202217, rodgers2024diamond}. For the hydrogen‐terminated (111) diamond surface, we studied a (\(1 \times 1\)) configuration, suggested to be more stable than the \((2 \times 1)\) \cite{tiwari2011calculated}. For each termination, the top and bottom surfaces were symmetrically passivated.  The geometry of all structures was optimized using DFT with the PBE functional \cite{perdew1996generalized}. We employed the self-consistent potential correction (SCPC) scheme to eliminate spurious electrostatic interactions from the compensating background under periodic boundary conditions for our charged diamond surfaces and to obtain reliable total energies and eigenvalues \cite{SCPC}. Our optimized models, shown in Fig.~\ref{fig:FIG1}, compare well with those previously reported in the literature \cite{kaviani2014proper, petrini2007theoretical,  stekolnikov2003tetramers}. 
To include surface defects (or dark spins) on the diamond surface, we used two atomistic motifs supported by experiments and theory \cite{stacey2019evidence, chou2023ab,sung2024identification}:  (i) An sp\(^3\) dangling bond (DB) on a (100) surface or step edge produced by incomplete passivation of a surface dimer, leaving one carbon under-coordinated. This creates a localized S = 1/2 defect whose stability depends on the local reconstruction and termination \cite{chou2023ab}. (ii) An sp\(^2\) defect generated by removing a surface carbon, after which neighboring atoms reconstruct to form a C=C unit and yield a more extended paramagnetic site \cite{stacey2019evidence}.  These two defect models offer a realistic representation of the structural variations on diamond surfaces that may impact the NV coherence properties.

\section{Results}
\subsection{Decoherence from Nuclear Spins as a function of surface termination}
\begin{figure}   
    \includegraphics[width=0.44\textwidth]{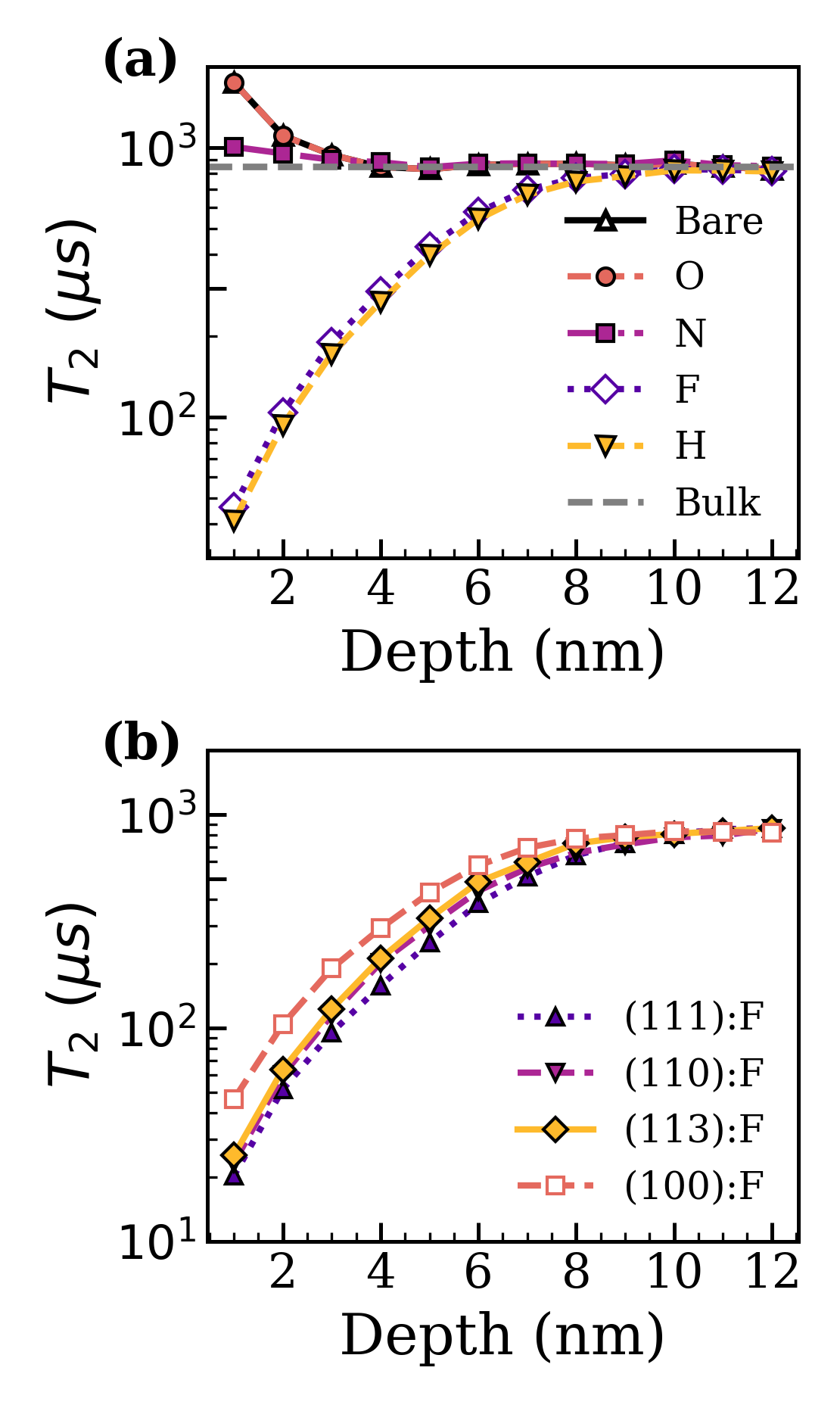}
    \caption{Hahn-Echo T$_2$ at various NV depth from diamond surfaces with various terminations. \textbf{(a)} Result for (100) surfaces terminated with O, N, F, H are compared with those obtained for NV in bulk diamond and in the presence of a bare surface (see text). \textbf{(b)} Hahn echo T$_2$ for Fluorine (F) terminations across various surface orientations.}
\label{fig:FIG2}
\end{figure}

 We start by investigating the decoherence of NV centers induced by nuclear spins and ligand terminations on (100) diamond surfaces, focusing on the effect of different atomic species and ligand functionalities on the computed Hahn-echo coherence time $(T_2)$.
 We vary the NV distance from the surface from 1 to 20 nm to identify the depths at which the NV coherence time approaches its bulk value.
As a computational reference, we constructed a hypothetical \textit{bare} diamond surface that is artificially saturated without any explicit ligands. While such a surface does not correspond to a physically realizable configuration, it allows us to identify a reference value for the NV decoherence time entirely due to the diamond lattice. In bulk diamond  with a natural isotopic concentration of \({}^{13}\)C nuclear spins (natural abundance 1.1\%, \(I = 1/2\)), the measured ensemble-averaged Hahn-echo coherence time (\(T_2\)) reaches 0.85~ms~\cite{maze2008electron, stanwix2010coherence}. We correctly reproduced this value in our CCE calculations for the bulk (Fig.~\ref{fig:FIG2}a). Notably, the bulk value is recovered for a \textit{bare} diamond surface at an NV depth of 4~nm, with NV centers near the \textit{bare} surface naturally exhibiting longer coherence times than those in the bulk, as  the number of contributing spins decreases as the NV approaches the surface. For example, in bulk diamond with 1.1\% \({}^{13}\)C, within a spherical region of radius  \(R \approx 5\)~nm, there are approximately 40{,}000 \({}^{13}\)C spins; this number decreases to about 20{,}000 for an NV center located 1~nm below the surface, leading to an increase in \(T_2\) by a factor of approximately 2 at 1~nm depth. 

As expected, for oxygen-terminated surfaces, the computed \(T_2\) values are essentially equivalent to those for the bare surface, reaching bulk values at an NV depth of 4~nm. The oxygen-termination introduces only a negligible additional spin bath, the spin-active $^{17}$O nuclei ( $ I=5/2$), with an extremely low natural abundance of $0.038\%$. Moreover, the inherent quadrupolar interactions of $^{17}$O suppress some nuclear flip-flop dynamics, minimizing its contribution to the NV decoherence beyond the baseline imposed by the $^{13}$C nuclear spin bath.  We also find that the specific oxygen termination, whether with ketone (C$=$O) or ether (C$-$O$-$C) groups, has no significant impact on $T_2$. 

Nitrogen-terminated surfaces induce only a slight reduction in \(T_2\) for NV centers at depths of 1~nm, with the coherence time converging to the bulk limit at 2~nm. In contrast, hydrogen- and fluorine-terminated surfaces generate substantial surface noise, resulting in a marked reduction in \(T_2\) over depths 1 to 12~nm. This degradation is attributed to the much larger nuclear gyromagnetic ratios of hydrogen (\(^{1}\)H) and fluorine (\(^{19}\)F), which are about 13 times that of nitrogen (\(^{14}\)N), even though these species are present at $\approx $ 99\% surface coverage.
We note that from our DFT calculations, the NV centers at a depth of 1~nm exhibit nonzero transverse zero‐field splitting (\(E\)), due to surface-induced symmetry reduction from \(C_{3v}\). However, the computed \(E\) values are on the order of 10~MHz, which is less than 1\% of the Zeeman splitting at 400~G ( 1120~MHz), and therefore they represent only a small perturbation that does not significantly affect the coherence time (See Appendix C).

Furthermore, we find that NV coherence depends on both the surface orientation and the spatial arrangement of the terminating nuclear spins. We computed \(T_2\) as a function of NV depth for fluorine-terminated diamond surfaces with (100), (111), (110), and (113) orientations (see Fig~\ref{fig:FIG2} b). For each crystal facet aligned to the NV quantization axis, we find that within the first 8 nm from the surface, \(T_2\) is sensitive to  surface orientation. For example, at 4~nm, $T^{(111)}_2=157~\mu s$, $T^{(110)}_2=202~\mu s$ , $T^{(113)}_2=211~\mu s$ and $T^{(100)}_2 =293~\mu s$. These differences decrease with increasing depth and become negligible beyond 8 nm. The variation in \(T_2\) arises from geometric factors that determine dipolar couplings and from differences in the surface nuclear spin lattice arrangement and density. Our results support the proposal that fluorine-terminated diamond surfaces can serve as experimentally accessible quantum simulators \cite{cai2013large}, and they show that NV coherence encodes the microscopic geometry of the underlying spin lattice. Detailed analyses of these orientation-dependent effects are provided in Appendix D.\par\vspace{-\parskip}

\subsection*{B. Decoherence from localized unpaired  electron spins at the surface }

\begin{figure*}[htbp!]
    \centering
    \includegraphics[width=0.8\textwidth]{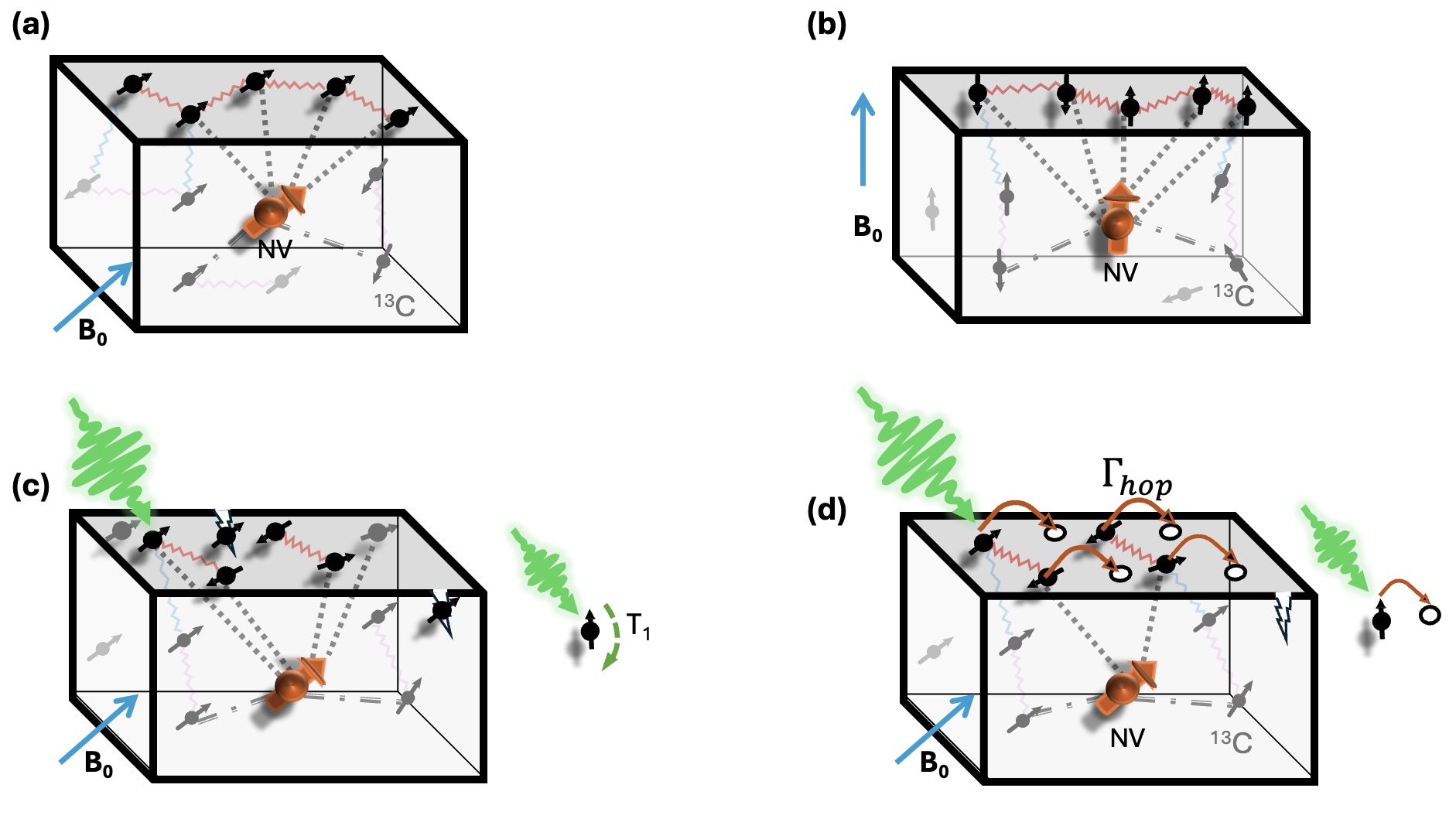}
    \caption{Representative models of electron spins on diamond surfaces. \textbf{(a-b)}. (100)  and (111) surfaces with surface electron spins (black arrows) aligned to the NV quantization axis; the NV center is shown in orange and nearby $^{13}C$ nuclear spins in gray. \textbf{(c)}. (100) surface electron in the presence of relaxation effects with relaxation time $T_1$ \textbf{(d)}. (100) surface electron spins subject to hopping, with rates $\Gamma_{hop}$; open circles indicate vacant sites. The direction of the external magnetic field $B_0$ is indicated in each case (see text).
    }
\label{fig:FIG3a}
\end{figure*}

We now turn to analyzing surface defects with an unpaired electron “dark” spins on the diamond surface (hereafter called surface electron spins or surface spins), which have been identified as a major source of NV center's  decoherence \cite{sangtawesin2019origins}. We consider two cases. (i) Long relaxation times $T_1$  $ > 100 ~ \mu s$, (or approaching infinite $T_1$)  that can result from suppressed flip-flop dynamics in disordered, interacting spin ensembles, as recently observed experimentally \cite{problongT1, GraesserPRR2023} or from cleaner and cryogenically cooled environments, where thermal fluctuations are suppressed and surface spins are more isolated . (ii) Finite relaxation times  $ T_1 \leq 100 ~\mu s$, which can result from thermally driven spin-lattice relaxation or from external microwave or radio frequency driving of surface spins \cite{bluvstein2019extending}, which accelerates spin flips (see Fig \ref{fig:FIG3a} (c)). 

\subsection*{I. Infinite relaxation time}

\begin{figure*}[htbp!]
    \centering
    \includegraphics[width=\textwidth]{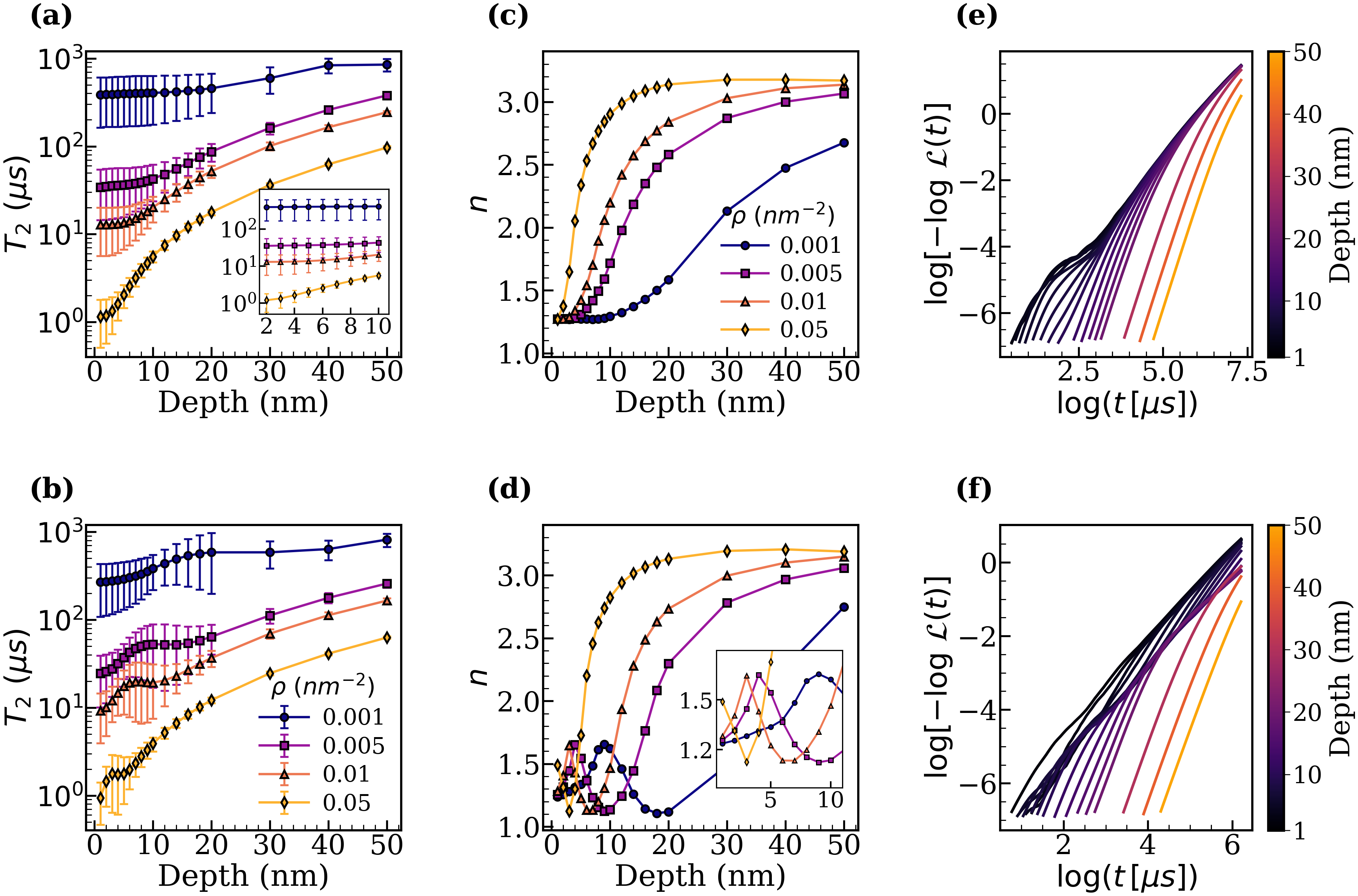}
    \caption{Computed Hahn echo T$_2$, for NV center for (100) and (111) surface against NV centers as a function of their depth from (100) \textbf{(a)} and (111) \textbf{(b)} surfaces, for various surface electron spin densities ($\rho$), as indicated in \textbf{(b)}.  Computed single stretched exponent of the Hahn-echo coherence function of NV centers as a function of their depth from (100) \textbf{(c)} and (111) \textbf{(d)} surfaces, for various surface electron spin densities ($\rho$), as indicated in \textbf{(c)}. Log-log plot of the Hahn-echo coherence signal showing the temporal transition of the stretching factors at various depths of the NV center for (100) \textbf{(e)} and (111) \textbf{(f)} surface. $\mathcal{L}(t)$ indicates the computed coherence signal. }
\label{fig:FIG3}
\end{figure*}

Experimental measurements indicate that the density of surface electron spins is in the range  \(\rho=0.001\text{–}0.05\,\mathrm{nm}^{-2}\) \cite{myers2014probing, dwyer2022probing, de2018density}. 
To quantify their effect, we model a two-dimensional ensemble of spins with dipolar interactions among themselves and with the NV center: each spin in the ensemble is localized at fixed positions (with randomly chosen configurations in different simulations), and has an effectively infinite relaxation time \(T_{1}\) (see Fig. \ref{fig:FIG3a}(a-b)).  Using CCE-4, we computed \(T_{2}\) as a function of depth $d_{NV}$ from the surface (\(1\le d_{NV}\le50\)\,nm), averaging over approximately $1500$ independent configurations.

In Fig.\ref{fig:FIG3}(c)  and  \ref{fig:FIG3}(d), we show the computed Hahn-echo \(T_{2}\) as a function of the NV depth for the (100) and (111) surface orientations at spin densities \(\rho=0.001,\,0.005,\,0.01,\) and \(0.05\,\mathrm{nm}^{-2}\). Compared to the decoherence induced only by nuclear spins, surface electron spins reduce $T_2$ by up to two orders of magnitude at shallow depths, depending on the \(\rho\), and on the surface orientation. On the (100) surface, where the NV axis is tilted by $54.7^{\circ}$ from the surface normal, \(T_{2}\) increases monotonically with depth; in contrast, on the (111) surface, where the NV axis lies parallel to the surface, $T_2$ is non-monotonic with subtle plateaus and dips that reflect geometric modulation of the coupling between the NV and the surface spin bath.
For both surface orientations, at shallow depths \(d_{NV}\lesssim5\)\,nm  the statistical spread of the fitted $T_2$ values across configurations can reach up to \(\sim50\%\) of the mean, due to the small number of proximal surface spins, which vary significantly from one configuration to another. These variations decrease with depth as coupling between the NV and surface spins weakens and the effective number of contributing spins grows, yielding small error bars.
For the most dilute bath $( \rho=0.001\,\mathrm{nm}^{-2})$ on the (100) surface, the mean $T_2$ and its standard deviation remain essentially unchanged in the range  (\(1\le d_{NV}\leq 10\)\,nm), rising only from roughly \(390\)\,µs  to \(400\)\,µs. By contrast, at higher surface spin densities $T_2$ exhibits a pronounced depth dependence, most notably  for \(\rho=0.05\,\mathrm{nm}^{-2}\), $T_2$ increases from \(\sim1 \ \mu s\)\, at 1 nm to \(\sim20 \ \mu s\)\, at 10 nm (See inset of Fig. \ref{fig:FIG3}(a)).

To further characterize the noise generated by surface electron spins in different surface orientations and NV depths, we computed the stretched exponent \(n\) of the Hahn-echo decay as a function of $d_{NV}$ and spin densities $\rho$.
As shown in Figs. \ref{fig:FIG3}(c) and \ref{fig:FIG3}(d), the nature of surface spin induced decoherence varies with increasing NV depth. For the (100) surface orientation, when $d_{NV}$ is much smaller than the characteristic interspin distance of the bath, \(d_{\rm NV}\ll d_{ij}=1/\sqrt{\rho}\), the stretched exponent remains close to \(n\approx1.3\), indicating nearly exponential decay characteristic of a fast‐fluctuating bath (\(n<2\)) \cite{de2009electron}. As \(d_{\rm NV}\) approaches \(d_{ij}\), \(n\) increases sharply, making a crossover to a quasi‐static noise regime. In the limit \(d_{\rm NV}\gg d_{ij}\), \(n\) saturates near 3, consistent with a compressed‐exponential decay caused by a spin environment that is effectively static on the timescale of the Hahn-echo sequence. The depth at which the crossover between fast fluctuating and static regimes occurs depends strongly on the spin density: higher densities lead to a transition at shallower depths. For example, at $\rho=0.05~nm^{-2}$, the stretched exponent $n$, begins to rise for  \(d_{\rm NV} = 1 ~nm\), whereas for $\rho=0.001~nm^{-2}$, $n$ remains nearly constant within $10~nm$ from the surface. These findings indicate that increasing the spin density leads to a faster variation of the bath dynamics with depth; in dilute regimes the transition from fast to quasi-static noise is more gradual and occurs deeper below the surface.

At variance from the (100) surface,  in the presence of a (111) surface the stretch exponent exhibits a non-monotonic depth dependence: it rises to a local maximum at \(d_{\rm NV}\approx d_{ij}/4\), then falls sharply to a minimum near \(d_{\rm NV}\approx d_{ij}/2\), before increasing again at larger depths, where it follows the same monotonic trend as found for the (100) surface. This non-monotonicity is also evident in the Hahn-echo coherence traces plotted on a log–log scale, which reveal temporal crossovers in the effective stretching behavior at different NV depths on the (111) surface (see Fig.\ref{fig:FIG3}(f)). 

\begin{figure}
    \centering
    \includegraphics[width=0.5\textwidth]{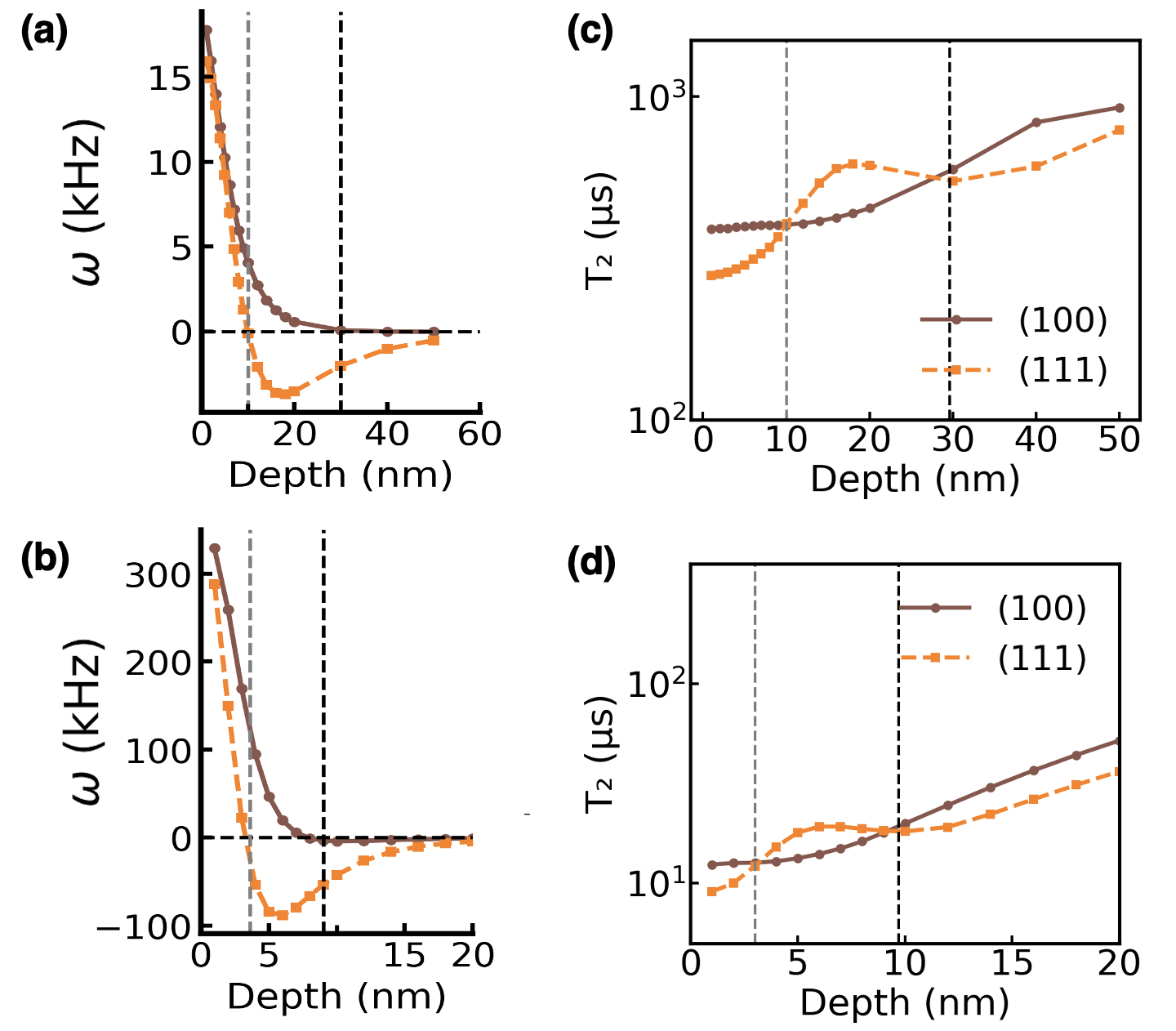}
\caption{Precession frequency \(\omega\) of the most strongly coupled surface pseudo-spin as a function of NV center's depth for surface‐spin densities \(\rho = 0.001\rm\ nm^{-2}\) \textbf{(a)} and \(\rho=0.01\rm\ nm^{-2}\) \textbf{(b)}. Computed Hahn–echo coherence time \(T_{2}\) as a function of depth at \(\rho=0.001\rm\ nm^{-2}\) \textbf{(c)} and \(\rho=0.01\rm\ nm^{-2}\) \textbf{(d)}.  Solid brown lines indicate the results for (100) surface, and dashed orange lines for (111) surface.  Vertical dashed lines indicate the characteristic depths at which the (111) surface exhibits its coherence “sweet spot.” (see text)}
\label{fig:pseudo}
\end{figure}

Interestingly, the observed non-monotonic variations in both the stretched exponent and Hahn-echo $T_2$ indicate preferred distances and geometric configurations leading the (111) face to have a higher $T_2$ compared to the (100) case. As shown in Fig. \ref{fig:pseudo} (b) and (d), for the same spin density, at some  characteristic depths, the $T_2$ of the NV is about 40 \%  higher in the presence of a (111) surface than for the (100) face, in spite of the favorable magic angle of the NV on the (100) surface. 
To understand in detail how this complex dynamics emerges, we considered a regime where only pairwise flip-flop interactions occur and their exchanges drive the decoherence process. By focusing on a single pair \textit{i}, \textit{j}, 
 we map its evolution onto a two-level  pseudo-spin model \cite{yao2006theory} $
  \Uparrow\!\bigl(\ket{\uparrow\downarrow}\bigr)\;\leftrightarrow\;
  \Downarrow\!\bigl(\ket{\downarrow\uparrow}\bigr)\,$, given by the Hamiltonian: 
\begin{equation}
    H^{\text{pseudo}} = \frac{\omega(t)}{2}\, \hat{\sigma}_z + \frac{\delta}{2} \,\hat{\sigma}_x.
\end{equation}
Here $
\omega(t)=\bigl|A^i_{\parallel}-A^j_{\parallel}\bigr|$ is the pseudo-spin frequency, set by the difference in dipolar couplings of surface spins \(i\) and \(j\) to the NV center.  The axial coupling of spin \(i\) is \(A^i_{\parallel} = \frac{\hbar\,\gamma_i\,\gamma_{\mathrm{NV}}}{2\,\lvert\mathbf d_{\mathrm{NV},i}\rvert^3}(3\cos^2\!\theta - 1)\), where \(\mathbf d_{\mathrm{NV},i}\) is the vector from the NV center to spin \(i\), and \(\theta\) is the angle between \(\mathbf d_{\mathrm{NV},i}\) and the external magnetic field. Similarly, the flip‐flop coupling between spins \(i\) and \(j\) is \(
\delta
= \frac{\hbar\,\gamma_i\,\gamma_j}{2\,\lvert\mathbf d_{ij}\rvert^3}
\bigl(3\cos^2\!\phi-1\bigr),
\)  with \(\mathbf d_{ij}\) connecting spins \(i\) and \(j\) and \(\phi\) its angle to the field.

The total coherence function $L(t)$ can be obtained analytically \cite{seo2016quantum} as a product of the contributions of each spin pair :

\begin{equation}
   L(t) = \prod_{ij}\bigg[1-\kappa\cdot \text{sin}^2(\sqrt{\omega^2 + \delta^2}\frac{t}{2}) \text{sin}^2(\delta^2\frac{t}{2}) \bigg] ,
\end{equation}

where, $\kappa = \frac{\omega^2}{\omega^2 + \delta^2}$. 
We examined 200 spin pairs contributing to decoherence,  in close proximity to the NV center. Specifically, we analyzed how the pseudospin frequency $\omega (t)$ varies with NV depth for the (100) and (111) surfaces (Fig. \ref{fig:pseudo} (a-d)) at two surface densities, 0.001 nm$^{-2}$ and 0.01 nm$^{-2}$.  As $d_{\text{NV}}$ increases, the frequency of the pseudospin $\omega(t)$ on the (100) face decays to zero at  \(d_{\rm NV}\approx d_{ij}\). On the (111) face, $\omega (t)$ changes sign at \(d_{\rm NV}\approx d_{ij}/4\), reaches a local minimum at  \(d_{\rm NV}\approx d_{ij}/2\), and then approaches zero again as \(d_{\rm NV}\approx d_{ij}\). When $\omega(t) = 0$, the two spin pairs have identical dipolar couplings to the NV center and their contribution to the coherence function becomes negligible, leading to the T$_2$ crossover observed on the (111) surface. The depth at which $\omega (t)$ attains its minimum corresponds to a maximum in T$_2$, which then declines as $\omega (t)$ approaches zero.

The distinct evolution of $\omega (t)$ with surface orientation underlies the non-monotonic dependence of T$_2$ and of the stretched exponent $n$ on the NV depth. These results show that both the magnitude and depth dependence of the NV-center decoherence are dictated by the interplay between surface spin density and crystallographic orientation and NV depth. Hence, they highlight pathways to optimize NV coherence by choosing facet geometry and orientation.

Finally, we also assessed how temperature modifies these dipolar interactions: we found that cooling into the millikelvin range (with thresholds set by $\rho$ and $d_{\rm NV}$) the bath can be can thermally polarized leading to the quenching of flip-flop mediated noise, and restoring bulk-limited $T_2$ (See Appendix E ).

\subsection*{II. Finite relaxation times}

\begin{figure*}[htbp!]
    \centering
    \includegraphics[width=\textwidth]{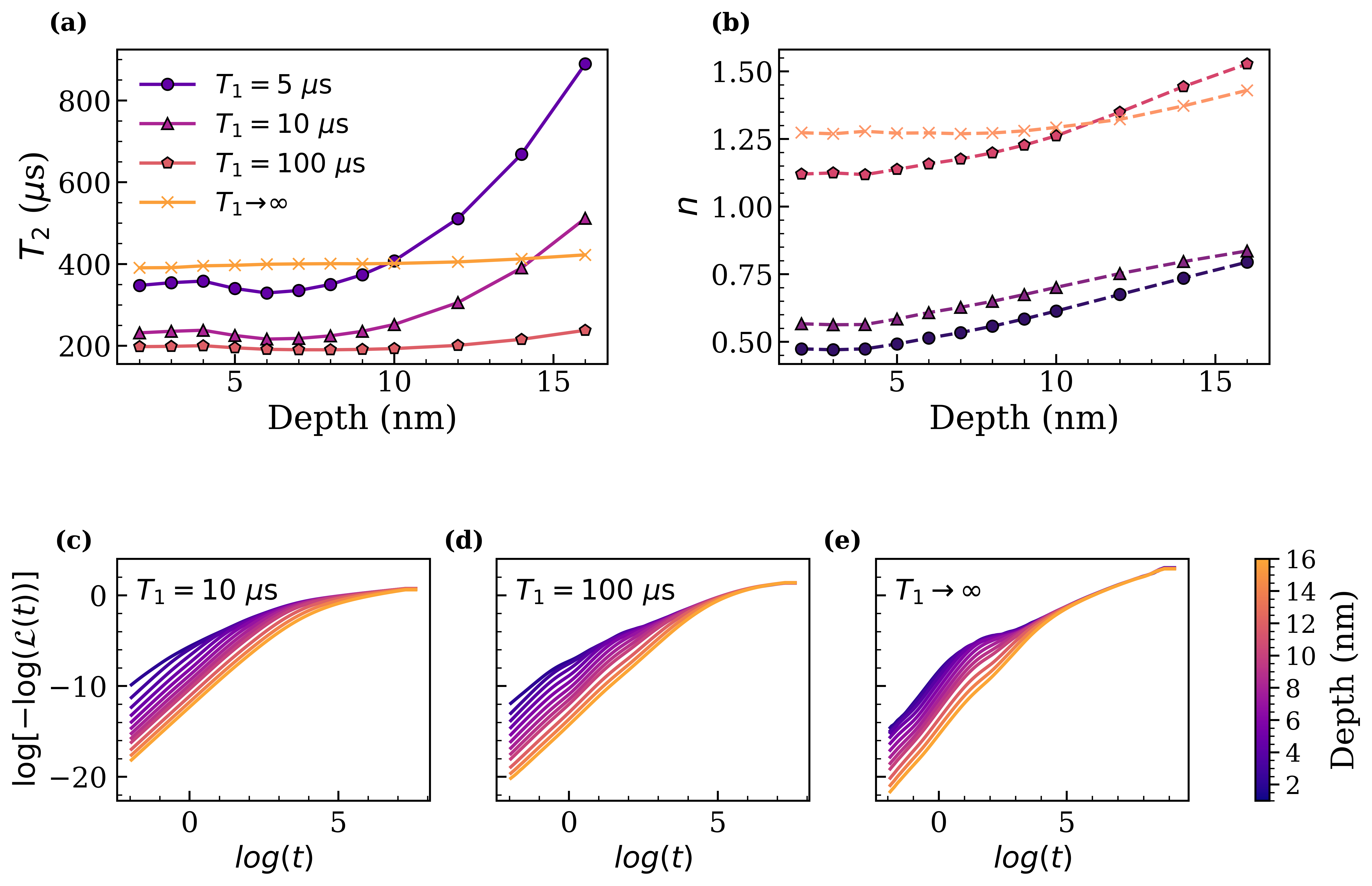}
    \caption{: \textbf{(a)} Hahn‐echo coherence time $T_2$ computed for a dilute surface‐spin bath ($\rho=0.001 \ nm^{-2}$), as a function of NV center's depth  and surface-spin relaxation time $T_1$. \textbf{(b)}. Extracted stretched exponent $n$ from the coherence decay as a function of depth for the same $T_1$ values shown in \textbf{(a)}. \textbf{(c-e)} Log–log plots of the Hahn‐echo decay at representative depths, showing how the upward curvature and instantaneous exponent evolve with $T_1$. }
\label{fig:FIG7}
\end{figure*}

Next we consider the case of finite relaxation time: unpaired electron spins on the diamond surface do exhibit finite spin relaxation or depolarization (T$_1$) time, making them useful “quantum reporters”  \cite{sushkov2014magnetic, espinos2024enhancing, de2018density}.
 The surface spin relaxation has been measured experimentally and has been found to be at least  $20 \ \mu s $ \cite{T1_nspins}. A finite relaxation rate of surface spins $\Gamma_i=1/(2\pi \ T_1$) generates stochastic flips that, together with coherent dipolar flip-flops between surface spins \(\delta
= \frac{\hbar\,\gamma_i\,\gamma_j}{2\,\lvert\mathbf d_{ij}\rvert^3}
\bigl(3\cos^2\!\phi-1\bigr),
\) shorten the bath correlation time $\tau_c$ and reshape the noise spectrum. These processes limit the NV center's coherence compared to the case of infinite-relaxation time limit (T$_1 \rightarrow \infty $) discussed in the previous section. To quantify the effect of finite T$_1$, we considered a dilute surface-spin bath with surface density $\rho=0.001\ \mathrm{nm}^{-2}$ and with $T_1$ varying from $5$ to $100 \ \mu s$ (Fig.~\ref{fig:FIG7}(a)). Using the ME-CCE formalism \cite{onizhuk2024understanding}, we modeled a dissipative spin bath as described in Eq.(\ref{eq:lindblad}), with  $\Gamma_{NV}=0$ , all inter-spin depolarization rates $\Gamma_{ij}=0$ and  $\Gamma_i=1/(2\pi \ T_1)$, and we computed the $T_2$ and stretched exponents as function of the NV depth ($d_{NV} = 1$–16 nm).
Fig.~\ref{fig:FIG7}(b) shows that the NV coherence time varies non monotonically with depth and surface spin $T_1$. At shallow depths (1–4 nm), the NV–surface-spin coupling strength ($A^i_{\parallel}$ ) is much larger than  both  $ \Gamma_i$, and  the mean surface-spin-bath coupling $\delta_{0}=\frac{\hbar\,\gamma_i\,\gamma_j}{2\,\lvert\mathbf d_{0}\rvert^3}
\bigl(3\cos^2\!\phi-1\bigr)
$ with \(\mathbf d_{0}=1/\sqrt{\rho}\) ; in this under damped regime, spin flips are rare on the echo timescale but, when they occur, they induce sudden spectral jumps (spectral diffusion) \cite{regimes2}. As a result, T$_2$ is reduced and the stretch exponent falls below unity, in contrast to the case of the undamped limit ($\Gamma_i=0$, $T_1 \rightarrow \infty)$, indicating a broad distribution of bath correlation times. 
At intermediate depths (for example, 5–9 nm and $T_1=5 \ \mu s$), $A^i_{\parallel}$ becomes comparable to $ \Gamma_i$, and still exceeds $\delta_{0}$. In this case, we observe a pronounced minimum in $T_2$ despite the increased NV depth. This critical damping arises because surface-spin flips mix with the coherent dipolar flip-flops $\delta$; they thus maximize fluctuations in the NV’s local magnetic field on the echo timescale, thereby minimizing $T_2$. At larger depths, where both $A^i_{\parallel}$ and $\delta_{0}$ are much smaller than $\Gamma_i$, surface spin flips are so fast that they effectively decouple the bath from the NV center and lead to a motional-narrowing regime with drastically enhanced coherence times \cite{regimes}. This behavior persists at NV depths exceeding the surface spin nearest neighbors coupling distances, where the bath dynamics are characterized by a single correlation time $\tau_c=T_1$ ; accordingly, the stretch exponents saturates at $n=1$. 
The influence of finite surface spin relaxation on NV centers' coherence is also evident in  Figs.~\ref{fig:FIG7}(c–e). As \(T_{1}\) increases, spin flips become slower and coherent interactions dominate; in this case we observe a single power-law decay with a uniform slope, reflecting one dominant, long-lived correlation process. In contrast, when $T_1$ is short relative to the Hahn-echo evolution time, the pronounced upward curvature indicates that the NV center exhibits a broad distribution of correlation times~\cite{davis2023probing}. At any fixed \(T_{1}\), deeper NV centers show a larger exponent \(n\) and reduced curvature. In the \(T_{1}\rightarrow\infty\) limit, increasing depth merely shifts the decay time without affecting its slope or curvature.


\begin{figure*}[htbp!]
    \centering
    \includegraphics[width=0.9\textwidth]{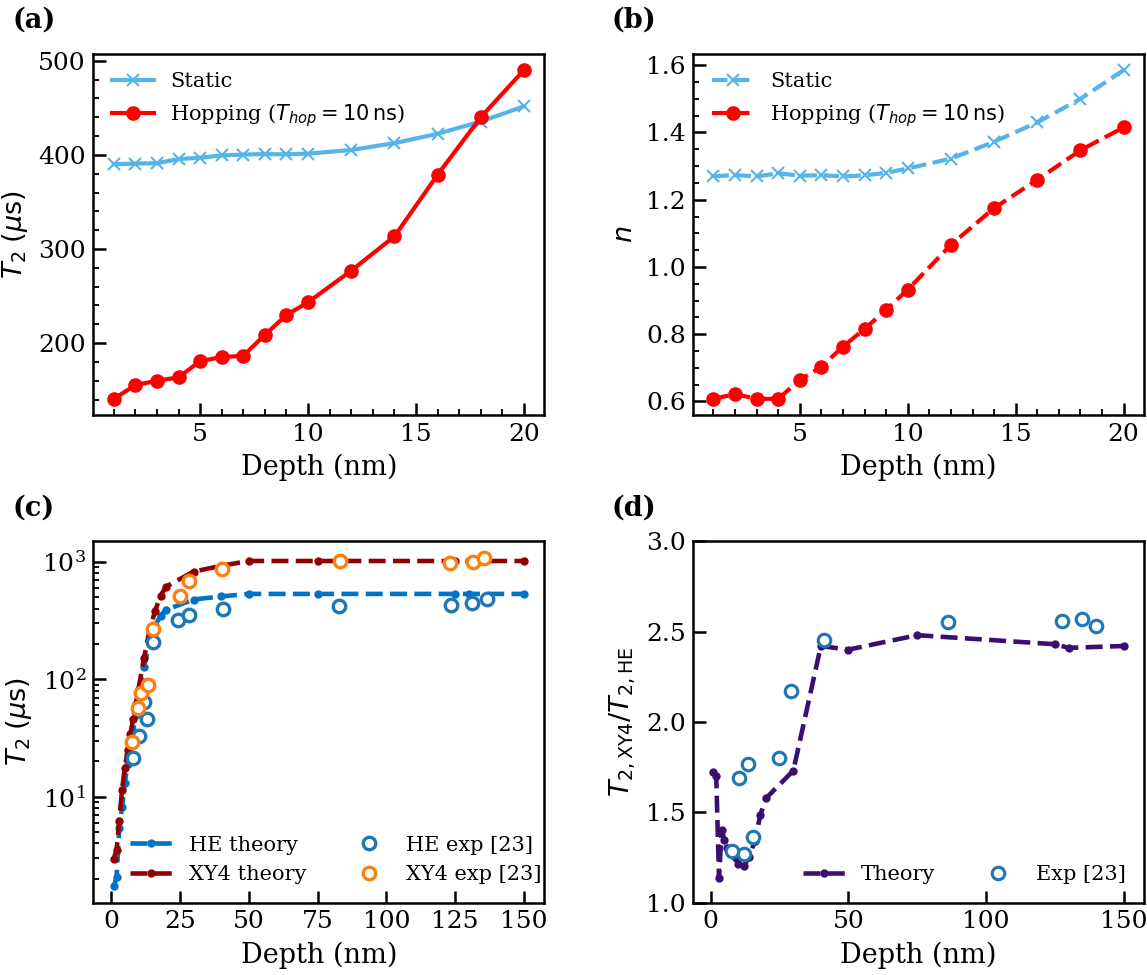}
    \caption{\textbf{(a)} Computed Hahn-echo coherence time $T_2$ as a function of NV centers' depth static (blue line) and hopping (red line) surface spin models, at a spin density $\rho=0.001~nm^{-2},$ showing distinct depth-dependent scaling behavior. \textbf{(b)} Stretched-exponent fit parameter $n$ as a function of NV depth, for static and hopping surface spin models.  $T_{hop}$ denotes the characteristic surface-spin hopping time \textbf{(c)} Computed Hahn-echo and XY4 $T_2$
  times (dashed lines) at a spin density $\rho=0.004~nm^{-2},$ compared with experimental data (open circles) from Ref.\cite{myers2014probing}  \textbf{(d)} Ratio of XY4 to Hahn-echo coherence times as a function of NV centers' depth, showing good agreement between theory (dashed line) and experiment (open circles), and indicating the influence of high-frequency bath noise at shallow depths.
  }
    \label{fig:FIG9}
\end{figure*} 
Because of the non-monotonic behavior of $T_2$ with depth at finite $T_1$, one can exploit the motional narrowing induced by fast relaxation of surface spins to cancel their contribution to NV coherence. As shown in Appendix F, for very short surface-spin $T_1 = 1~\mu\text{s}$, the NV coherence becomes completely bulk-limited, showing no influence from surface spins.

The results of this section highlight how the interplay between the NV-bath coupling strength, surface spin relaxation rate, and surface spin bath dipolar interactions lead to different NV center’s decoherence regimes, that range from under-damped spectral diffusion to motional narrowing.  Hence our calculations provide a microscopic interpretation of the improvement in coherence observed experimentally under stochastic bath driving \cite{bluvstein2019extending}.

\subsection*{C. Decoherence from  unpaired electron spins hopping between sites}

Recent experimental studies have shown that, after certain surface treatments, some surface spins become mobile under green‑laser illumination\cite{sushkov2014magnetic, dwyer2022probing}. This laser‑induced spatial reconfiguration generates both magnetic and electric fluctuations that contribute to the dephasing of the NV center \cite{candido2024theory, dwyer2022probing}. To model this behavior, we incorporated time-dependent, in-sequence hopping processes into our framework. Specifically, we described the bath dynamics through the hopping motion of trapped unpaired electrons to nearby acceptor states, as illustrated in Fig. \ref{fig:FIG3a}(d).

In our approach, spins are allowed to hop to a neighboring acceptor state or be trapped only when their mutual separation falls below a critical distance $r_c$. We used experimentally supported values of the hopping timescale \(T_{\mathrm{hop}} \approx 10\) ns and $r_c=5$ nm \cite{choi2017depolarization, chou2018} and simulated the spin dynamics with the MECCE framework, as described by Eqs. (10) and (11).  The hopping dynamics is governed solely by pairwise hopping rates, given by  $\Gamma_{ij} = 1 /(2\pi \ T_{hop} ) \   exp \ ({-r/r_c})$; the corresponding pairwise Lindblad jump operators that define these processes are detailed in Appendix G.

Our results show that the hopping model introduces a significant monotonic depth dependence in the NV center’s decoherence time  $T_2$, distinguishing it from other surface spin models that do not include hopping. Notably, the stretched exponent characterizing the decay starts at 2/3 and increases steadily with depth. To validate our model, we compared our simulation results to the experimental data from Myers et al.\cite{myers2014probing}, who studied NV centers in a (100)-oriented, isotopically purified CVD-grown diamond using nitrogen delta-doping to control depth. Their measurements reveal a strong suppression of both $T_2^{HE}$ and $T_2^{XY4}$ for NVs shallower than $\approx 25~nm$, with a corresponding decrease in the dynamical decoupling efficiency $\lambda=T_2^{XY4}/T_2^{HE}$, which drops from 2.5 in the bulk-like regime to values as low as $1.2$ near the surface. We find that including in-sequence surface spin hopping is essential to reproduce the trends observed experimentally, particularly the reduced coherence enhancement at shallow depths (see Fig. \ref{fig:FIG9}(c-d)). By incorporating physically motivated, time-dependent hopping processes into the ME-CCE framework, our model reconciles the key discrepancies between static theoretical predictions and experimental observations. Our results highlight the critical role of environment-induced, non-static disorder in realistic modeling of surface spin noise affecting shallow NV centers.

\section{Conclusion}
In summary, we developed and validated a comprehensive theoretical framework that captures the full complexity of NV-center decoherence at diamond surfaces. By integrating DFT-derived atomistic surface models with coherence time calculations carried out with the CCE approach, we quantified the distinct contributions of nuclear spin baths, static surface electrons, spin relaxation, and hopping dynamics across a wide range of conditions. We considered varied depths of the NV sensor, different surface terminations, crystallographic orientations, and temperatures. Our results reveal that hydrogen and fluorine terminations significantly reduce coherence times at shallow depths, while oxygen and nitrogen terminations preserve bulk-like $T_2$. We showed that the experimentally observed transition from exponential to compressed-exponential decay can be predicted by the ratio of the NV depth from the surface to the surface spin-spin distances. Interestingly, the choice of crystallographic orientation further enables coherence “sweet spots”, particularly on the (111) surface, where symmetry-induced dipolar coupling cancellation suppresses surface spin noise. Importantly, we demonstrated that surface electron spin relaxation can induce motional narrowing, restoring bulk-like coherence times even at shallow depths. Crucially, we found that including surface spin hopping is necessary to reproduce the experimentally observed depth dependence of $T_2$, thus establishing hopping as a dominant decoherence channel and confirming the NV center's sensitivity to itinerant surface carrier dynamics. 

The insights obtained in our work, together with our interpretation of experiments,  provide actionable design principles for optimizing the coherence properties of NV centers close to diamond surfaces; these principles include: the use of low-gyromagnetic-ratio nuclear spins for surface functionalization; an appropriate choice of crystallographic symmetry and of the depth of the NV center; and  specific ways to drive the surface electron spins that lead to motional narrowing. The predictive models proposed in our work enable effective strategies to mitigate decoherence, advancing NV-based quantum sensing.

\vspace*{1em}
\textit{Codes used.}
The master-equation CCE (ME-CCE) implementation, including stochastic hopping dynamics, is available as a part of the PyCCE package  \cite{onizhuk2021pycce}.

\vspace*{1.5em}
\begin{acknowledgments}
We thank Xiaofei Yu, Yuxin Wang and Peter Maurer for helpful discussions. 
This work was supported by the NSF QuBBE Quantum Leap Challenge Institute
and was carried out with the use of resources provided by the University of Chicago’s Research Computing Center, (RCC). The PyCCE code developement was supported by MICCoM, a DOE/BES funded Computational science center.

\end{acknowledgments}

\newpage
\appendix
\setcounter{section}{0}

\counterwithin{equation}{section} 



\renewcommand{\thesection}{\Alph{section}}
\renewcommand{\thesubsection}{\thesection.\roman{subsection}}

\makeatletter
\renewcommand{\p@subsection}{}
\makeatother
\FloatBarrier 
\section{Spin Coupling Tensors}
We used density-functional theory (DFT) to compute spin-coupling tensors for diamond surfaces with various terminations, considering NV centers positioned $ 10–20\AA$ below the surface; calculations employed a plane-wave basis set with a 80 Ry kinetic-energy cutoff. For the Zero-field splitting (ZFS) tensors  calculations we used projector-augmented-wave (PAW) pseudopotentials \cite{dal2014pseudopotentials}, while for the hyperfine interactions tensors we used gauge including projector augmented wave (GIPAW) \cite{Mauri} pseudopotentials.

\subsection{Zero-Field Splitting Tensors}

The components of the zero-field splitting (ZFS) tensor were approximated by considering only spin–spin
interactions, with calculations carried out using a GPU-accelerated version
of the \textsc{PyZFS} package \cite{ma2020pyzfs}. The spin–spin ZFS tensor was evaluated
from Kohn–Sham wavefunctions as
\begin{equation}
\label{eq:Dtensor}
\begin{aligned}
D_{ij} &=
  \frac{\mu_0}{4\pi}\,
  \frac{(\gamma_e\hbar)^2}{2S(2S-1)}
  \sum_{a<b}^{\mathrm{occ}} \chi_{ab}\; \\[2pt]
&\quad
  \big\langle \Phi_{ab} \big|
  \frac{ R_{ab}^2\,\delta_{ij}
        - 3\,R_{ab,i} R_{ab,j} }
       { R_{ab}^5 }
  \big| \Phi_{ab} \big\rangle ,
\end{aligned}
\end{equation}
where the summation is over all pairs of occupied Kohn–Sham orbitals.  
Here, $\chi_{ab}=\pm 1$ for parallel and antiparallel spins, respectively, and 
$\Phi_{ab}(\mathbf r,\mathbf r')$ is the antisymmetrized two-electron wavefunction,
\[
\Phi_{ab}(\mathbf r,\mathbf r')
=\tfrac{1}{\sqrt{2}}\big[\phi_a(\mathbf r)\phi_b(\mathbf r')
-\phi_a(\mathbf r')\phi_b(\mathbf r)\big].
\]
We define $\mathbf R_{ab}=\mathbf r-\mathbf r'$, with $R_{ab}=|\mathbf R_{ab}|$
and Cartesian components $R_{ab,i}$ $(i,j\in\{x,y,z\})$. Here $\mu_0$ denotes the vacuum permeability, and  $\gamma_e$ is the electron spin gyromagnetic ratio.

The scalar ZFS parameters are
\begin{equation}
    D=\tfrac{3}{2}D_{zz}, \qquad 
    E=\tfrac{1}{2}\,(D_{xx}-D_{yy}),
\end{equation}
corresponding to the axial and transverse anisotropy, respectively.

\subsection{Hyperfine Tensors}

The hyperfine tensor for a nuclear spin consists of two contributions:  
(i) an isotropic Fermi-contact term $A_{\mathrm{iso}}$, determined by the electron spin
density at the nuclear site, and (ii) a traceless anisotropic dipolar term
$A^{\mathrm{dip}}_{ij}$, arising from electron–nuclear magnetic dipole–dipole
interaction. The decomposition reads
\begin{equation}
\label{eq:A_decomp}
A_{ij}=A_{\mathrm{iso}}\,\delta_{ij}+A^{\mathrm{dip}}_{ij}, \qquad
A_{\mathrm{iso}}=\tfrac{1}{3}\,\mathrm{Tr}\,\mathsf{A}, \qquad
\mathrm{Tr}\,\mathsf{A}^{\mathrm{dip}}=0,
\end{equation}
where $\mathsf A$ is the $3\times 3$ hyperfine tensor.

The individual terms can be written in terms of the spin density $n_s(\mathbf r)$ as
\begin{equation}
\label{eq:A_iso}
A_{\mathrm{iso}}
= -\,\frac{\mu_0 \hbar^{2}}{3S}\,\gamma_e\,\gamma_N\, n_s(\mathbf r_N),
\end{equation}
\begin{equation}
\label{eq:A_dip_cont_gamma}
A^{\mathrm{dip}}_{ij}
= -\,\frac{\mu_0\,\hbar^{2}}{8\pi S}\,\gamma_e\,\gamma_n
\int \frac{R^{2}\delta_{ij}-3R_i R_j}{R^{5}}\,
n_s(\mathbf r)\,\mathrm d^{3}\mathbf r .
\end{equation}

with $\mathbf r_N$ the nuclear position and $\mathbf R=\mathbf r-\mathbf r_N$
($R=|\mathbf R|$, $R_i$ its Cartesian components).
Here $\gamma_N$ is the nuclear spin gyromagnetic ratio.
Hyperfine tensors were computed using two distance-dependent approaches relative to
the NV center. For nuclei within $20~\text{\AA}$, $\mathsf{A}$ was obtained
from density-functional theory (DFT) using the PBE functional and the GIPAW module of
\textsc{Quantum~Espresso} \cite{giannozzi2009quantum}, including both isotropic and dipolar terms.
For nuclei at distances $\geq 20~\text{\AA}$, the point-dipole approximation was used,
where the NV electronic spin is modeled as a point magnetic dipole at the defect site.
In this limit the Fermi-contact contribution vanishes and
\begin{equation}
\label{eq:A_dip_point_gamma}
A^{\mathrm{dip}}_{ij}
= -\,\frac{\mu_0\,\hbar^{2}}{4\pi}\,\gamma_e\,\gamma_n\,
\frac{R^{2}\delta_{ij}-3R_i R_j}{R^{5}} .
\end{equation}

\subsection{Quadrupolar Tensors}

The quadrupole interaction depends on the electric field gradient (EFG) tensor
$\mathsf V$, with components $V_{ij}$. At a nuclear site $\mathbf R_N$, the EFG is
\begin{equation}
\begin{aligned}
V_{ij}(\mathbf R_N)
&= \frac{1}{4\pi\varepsilon_0}\bigg[
\sum_{K\ne N} Z_K \,
\frac{ \big(R^{(K)}\big)^{2}\delta_{ij} - 3\,R^{(K)}_{i} R^{(K)}_{j} }
     { \big(R^{(K)}\big)^{5} } \\[2pt]
&\quad
-\; \int n_e(\mathbf r)\,
\frac{ R^{2}\delta_{ij} - 3 R_{i} R_{j} }{R^{5}}\,
\mathrm d^{3}\mathbf r
\bigg],
\end{aligned}
\end{equation}
where $n_e(\mathbf r)$ is the (positive) electron density,  
$\mathbf R^{(K)}=\mathbf R_K-\mathbf R_N$ is the vector from nucleus $N$ to $K$,  
and $\mathbf R=\mathbf r-\mathbf R_N$.

The nuclear quadrupolar coupling tensor $\mathsf Q$ is
\begin{equation}
Q_{ij}= \frac{e\,Q}{2I(2I-1)\hbar}\, V_{ij},
\end{equation}
with $Q$ the nuclear quadrupole moment and $I$ the nuclear spin quantum number.  

We computed quadrupolar parameters from DFT for $^{14}$N ($I=1$) and $^{17}$O ($I=5/2$) surface terminations using the GIPAW module of \textsc{Quantum~Espresso}.

\subsection{Magnetic Dipole–Dipole Coupling Tensors (Point Dipole)}

Dipolar interactions between spins (nuclear–nuclear or electron–electron) were
modeled in the point-dipole approximation, treating each spin as a point magnetic
moment.

The pairwise coupling kernel is
\begin{align}
K_{ij}(\mathbf R)=\frac{R^2\delta_{ij}-3R_iR_j}{R^5},
\qquad \mathbf R=\mathbf r_2-\mathbf r_1 .
\end{align}

Thus, the coupling tensors are
\begin{align}
P^{(\mathrm{nuc})}_{ij}(N,M)
  &= \frac{\mu_0}{4\pi}\,\gamma_N\gamma_M\,\hbar^2\,
     K_{ij}(\mathbf R_{MN}), \\
P^{(\mathrm{elec})}_{ij}(a,b)
  &= \frac{\mu_0}{4\pi}\,(\gamma_e\hbar)^2\,
     K_{ij}(\mathbf R_{ab}),
\end{align}
with $\mathbf R_{MN}=\mathbf R_M-\mathbf R_N$ and $\mathbf R_{ab}=\mathbf r_b-\mathbf r_a$.

These tensors are symmetric and traceless, since $\mathrm{Tr}\,K=0$.
\begin{equation}
   \mathrm{Tr}\,P^{(\mathrm{nuc})}=\mathrm{Tr}\,P^{(\mathrm{elec})}=0 . 
\end{equation}


\subsection{Static Electric Field from Surface Charges}

The electric field entering the NV Hamiltonian 
($\mathcal E_z,\mathcal E_x,\mathcal E_y$ in the NV principal frame)
is obtained by integrating the Coulomb field of a uniformly charged circular
\emph{region} on the surface, centered at the NV position $\mathbf r_{\mathrm{NV}}$, with 
the relative permittivity of diamond being $\varepsilon_d\!\approx\!5.7$).

We consider a region of radius $a$, centered at $\mathbf r_c$, with unit surface normal $\hat{\mathbf n}$, and surface charge density $\sigma$ (C/m$^2$). Points belonging to the region are defined as:
$\mathbf r'(\boldsymbol\rho)=\mathbf r_c+\boldsymbol\rho$, 
where $\boldsymbol\rho$ is an in-plane displacement vector lying in the surface plane (i.e. $\boldsymbol\rho\cdot\hat{\mathbf n}=0$) with magnitude restricted to $|\boldsymbol\rho|\le a$. 

The electric field inside the diamond slab at $\mathbf r_{\mathrm{NV}}$ is
\begin{equation}
\mathbf E(\mathbf r_{\mathrm{NV}})
= \frac{\sigma}{\varepsilon_0\,(\varepsilon_d+1)}
\int_{|\boldsymbol\rho|\le a}
\frac{\mathbf r_{\mathrm{NV}}-\mathbf r'(\boldsymbol\rho)}
{\big|\mathbf r_{\mathrm{NV}}-\mathbf r'(\boldsymbol\rho)\big|^{3}}
\, d^2\boldsymbol\rho .
\end{equation}
where $\epsilon_0$ is the vacuum permittivity. 
 We assume the NV lies on the symmetry axis beneath the region center, and we 
define the signed depth
\begin{equation}
z \equiv \hat{\mathbf n}\!\cdot\!\big(\mathbf r_{\mathrm{NV}}-\mathbf r_c\big),
\end{equation}
so that the field is normal to the surface, $\mathbf E(\mathbf r_{\mathrm{NV}})=
E_\perp(\mathbf r_{\mathrm{NV}})\,\hat{\mathbf n}$, with magnitude
\begin{equation}
E_\perp(\mathbf r_{\mathrm{NV}})
= \frac{\sigma}{\varepsilon_0\,(\varepsilon_d+1)}
\left(1-\frac{z}{\sqrt{z^2+a^2}}\right).
\end{equation}

Projecting this vector field onto the NV axes
$\{\hat{\mathbf e}_x^{\rm NV},\hat{\mathbf e}_y^{\rm NV},\hat{\mathbf e}_z^{\rm NV}\}$ gives:
\[
\mathcal E_i
= \hat{\mathbf e}_i^{\rm NV}\!\cdot\!\mathbf E(\mathbf r_{\mathrm{NV}}),
\qquad i\in\{x,y,z\}.
\]
which, on the symmetry axis, reduces to
\begin{equation}
\mathcal E_i
= \big(\hat{\mathbf e}_i^{\rm NV}\!\cdot\!\hat{\mathbf n}\big)\,
E_\perp(\mathbf r_{\mathrm{NV}}).
\end{equation}

The coupling of this static electric field to the NV spin levels is mediated by the ground-state electric susceptibility parameters of the NV center, with $d_\parallel \approx 3.5\times 10^{-3},\text{Hz},(\text{V},\text{m}^{-1})^{-1}$ and $d_\perp \approx 0.165,\text{Hz},(\text{V},\text{m}^{-1})^{-1}$.

\section{Lindblad Jump Operators}

We consider a bath of dipolar-coupled spin-\(\tfrac{1}{2}\) {spins}  interacting with a Markovian environment (as in the main text), with dynamics governed by the Lindblad master equation (Eq.~5). The dissipative processes are written using the standard dissipator
\begin{equation}
\mathcal{D}[L](\hat\rho)\equiv L\,\hat\rho\,L^\dagger-\tfrac{1}{2}\{L^\dagger L,\hat\rho\}.
\end{equation}

\paragraph{Single-spin jumps.}
For each bath spin \(i\in\mathcal N\),
\begin{equation}
L_i^{+}=\hat S^{\,i}_{+},\qquad
L_i^{-}=\hat S^{\,i}_{-},
\end{equation}
with corresponding rates \(\Gamma_i^{+}\) and \(\Gamma_i^{-}\). Here \(\hat{\mathbf S}^{\,i}\) are spin-\(\tfrac{1}{2}\) operators acting on the \(i\)-th bath spin.

\paragraph{Pairwise (incoherent) spin-exchange jumps.}
Between distinct spins \(i\neq j\),
\begin{equation}
L_{ij}^{\uparrow}=\hat S^{\,i}_{+}\!\otimes\!\hat S^{\,j}_{-},\qquad
L_{ij}^{\downarrow}=\hat S^{\,i}_{-}\!\otimes\!\hat S^{\,j}_{+},
\end{equation}
with rates \(\Gamma_{ij}^{\uparrow}\) and \(\Gamma_{ij}^{\downarrow}\). \\
For each channel we take forward and backward rates to be equal:
\begin{equation}
\Gamma_i^{+}=\Gamma_i^{-}\equiv \Gamma_i,
\qquad
\Gamma_{ij}^{\uparrow}=\Gamma_{ij}^{\downarrow}\equiv \Gamma_{ij}.
\end{equation}

\paragraph{Projected evolution for the NV coherence.}
To obtain an equation of motion for the NV off-diagonal block, we define the partial inner-product
\begin{equation}
\hat\rho_{01}(t)\equiv \bra{0}\hat\rho(t)\ket{1},
\end{equation}
which is not a density matrix and whose evolution is not generated by a Lindblad superoperator \cite{onizhuk2024understanding}. Here, \(\{\ket{0},\ket{1}\}\) are NV basis states. Starting from the total Lindbladian and retaining only jump operators that act on the bath spins, since the effect of an independent Markovian dephasing bath on the NV center is trivial and factors out, we obtain:
\begin{equation}\label{eq:first_projected}
\begin{split}
\frac{d}{dt}\hat\rho_{01}(t)
=\;& -\frac{i}{\hbar}\,\bra{0}\,[\hat H,\hat\rho(t)]\,\ket{1} \\
&\; + \sum_{i\in\mathcal N}\Gamma_i\,\bra{0}\!\left(\mathcal D[L_i^{+}](\hat\rho(t))
      + \mathcal D[L_i^{-}](\hat\rho(t))\right)\!\ket{1} \\
&\; + \sum_{i<j}\Gamma_{ij}\,\bra{0}\!\left(\mathcal D[L_{ij}^{\uparrow}](\hat\rho(t))
      + \mathcal D[L_{ij}^{\downarrow}](\hat\rho(t))\right)\!\ket{1}.
\end{split}
\end{equation}

Here \(\mathcal N\) denotes the set of bath spins, \(\{L_i^{\pm}\}\) are single-spin bath jumps, and \(\{L_{ij}^{\uparrow,\downarrow}\}\) are pairwise bath jumps with the rates defined above.

\section{Effects of local strain and electric field}

We investigated how proximity to the diamond surface breaks the NV center’s bulk \(C_{3v}\) symmetry and generates a transverse splitting \(E\) of the zero-field triplet. From our DFT calculations of the zero-field splitting, we find \(E \approx 35~\mathrm{MHz}\) for fluorine-terminated surfaces at a depth of \(\sim 1~\mathrm{nm}\), decreasing to \(\sim 2~\mathrm{MHz}\) by \(\sim 1.7~\mathrm{nm}\) as the NV approaches the bulk environment. Such near-surface symmetry breaking within the first \(1\text{–}3~\mathrm{nm}\) from the surface produces a linear Stark response at the NV location and mixes the \(\lvert +1\rangle\) and \(\lvert -1\rangle\) states into \(\lvert \pm \rangle\), so that the transitions are \(\lvert 0\rangle \leftrightarrow \lvert \pm \rangle\).

Additionally, prior NV electrometry measurements on both oxygen- and hydrogen-terminated diamond surfaces in air reported built-in fields from surface acceptors/adsorbates averaging \(\epsilon_z \approx 300\text{–}430~\mathrm{kV/cm}\) at depths \(7\text{–}35~\mathrm{nm}\), with modeled maxima near the surface of \(\approx 1.6~\mathrm{MV/cm}\) \cite{kim2015decoherence, broadway2018spatial}. Using the NV electric-dipole coefficients \(d_{\perp}=0.017~\mathrm{kHz}/(\mathrm{V}/\mathrm{cm})\) and \(d_{\parallel}=0.00035~\mathrm{kHz}/(\mathrm{V}/\mathrm{cm})\), such fields lead to static Stark shifts of \(\approx 0.15\text{–}0.5~\mathrm{MHz}\). Hence, we modeled the combined action of transverse zero-field splitting and the local electric field accordingly, with the Hamiltonian:.

\begin{equation}
\begin{split}
H_{\rm NV} \;=\;& \gamma_{\rm NV} B_z S_z + \bigl(D - d_{\parallel}\,\epsilon_z\bigr) S_z^2 \\
&+ \bigl(E - d_{\perp}\,\epsilon_y\bigr)\,(S_x^2 - S_y^2) \;-\; d_{\perp}\,\epsilon_x\,(S_x S_y + S_y S_x),
\end{split}
\end{equation}
where \(\epsilon_{x,y,z}\) are the components of the local electric field at the NV (in the NV principal axes). The effective transverse mixing that enters the transition frequencies is
\begin{equation}
E_{\rm eff} =\; \sqrt{\bigl(E - d_{\perp}\,\epsilon_y\bigr)^2 + \bigl(d_{\perp}\,\epsilon_x\bigr)^2},
\end{equation}
i.e., the combined contribution of the intrinsic transverse ZFS \(E\) and the transverse Stark terms.

Diagonalizing the \(\lvert \pm 1\rangle\) subspace gives the two \(\lvert 0\rangle \leftrightarrow \lvert \pm \rangle\) transition frequencies
\begin{equation}
f_{\pm} \;=\; D - d_{\parallel}\,\epsilon_z \;\pm\; \sqrt{(\gamma_{\rm NV} B_z)^2 + E_{\rm eff}^2}.
\end{equation}
The corresponding magnetic slope is
\begin{equation}
\frac{\partial f_{\pm}}{\partial B_z} \;=\; \pm \frac{\gamma_{\rm NV}^2\,B_z}{\sqrt{(\gamma_{\rm NV} B_z)^2 + E_{\rm eff}^2}}.
\end{equation}

\begin{figure}[H]
    \centering
    \includegraphics[width=0.5\textwidth]{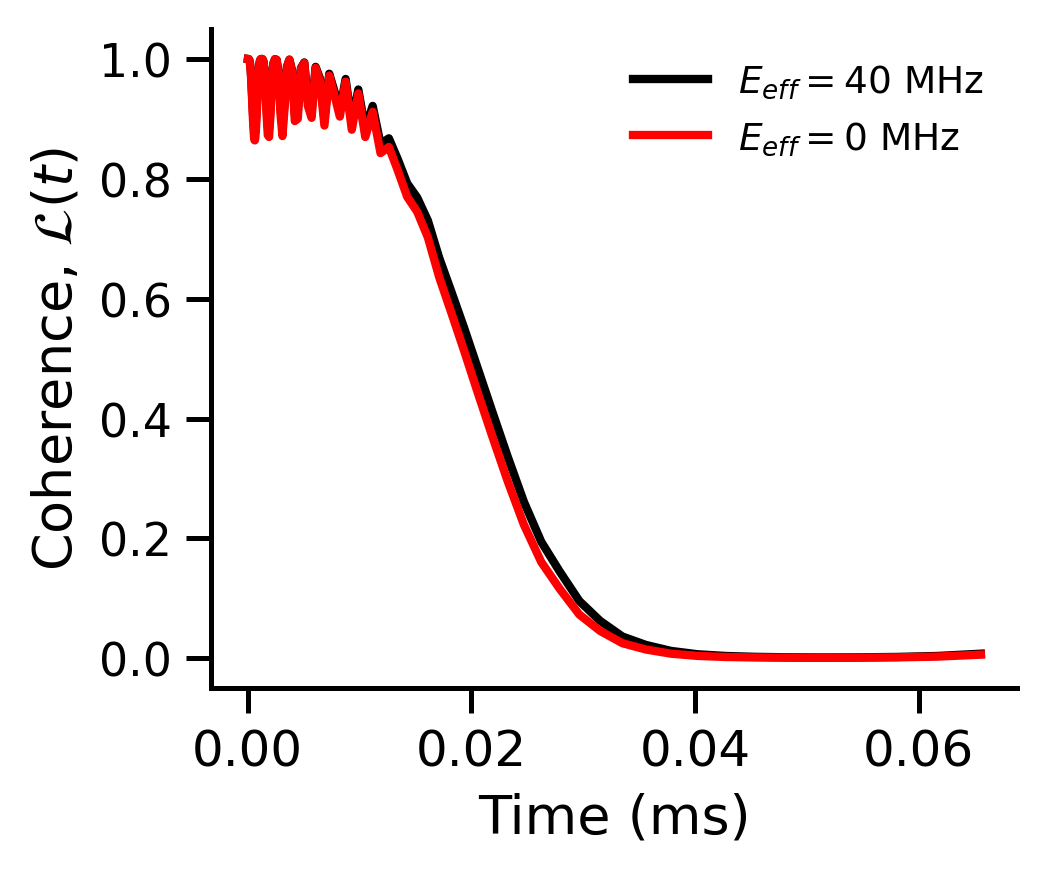}
    \caption{Coherence function $\mathcal{L}(t)$ of near-surface NV center as a function of an effective transverse Zero-field splitting and static electric field, $E_{eff}=0 \ MHz$ (red) and $E_{eff}=40 \ MHz $ (black) }
    \label{fig:Figure12}
\end{figure}

To illustrate the impact of transverse mixing, we consider two representative cases defined by the combined parameter \(E_{\rm eff}\): \(E_{\rm eff}=0\) and \(E_{\rm eff}=40~\mathrm{MHz}\). A comparison of the spectra and coherence times for these two cases (Fig.~\ref{fig:Figure12}) reveals only minor changes in the dephasing envelope, indicating that \(T_2\) is largely unaffected by a finite \(E_{\rm eff}\). For the field used here, \(B_z=400~\mathrm{G}\) (so \(\gamma_{\rm NV} B_z \approx 1120~\mathrm{MHz}\)), the fractional change of the magnetic slope is \(\approx \tfrac{1}{2}\,(E_{\rm eff}/\gamma_{\rm NV} B_z)^2\). Taking \(E_{\rm eff} = 40~\mathrm{MHz}\) gives \(\tfrac{1}{2}(40/1120)^2 \approx 0.064\%\). In contrast, when the field is sufficiently small ($B_z \leq E_{eff}/\gamma_{NV}$), transverse mixing suppresses the effect of magnetic noise near surfaces, which can result in a longer $T_2$ \cite{pershin2025}. 

Our simulations include only a static (DC) electric field at the NV arising from surface band bending, \cite{broadway2018spatial}. i.e, we treat $\epsilon_{x,y,z}$ as time-independent. Consequently, dephasing in our model is governed by magnetic surface noise. We note that electric noise from charge fluctuations can contribute to decoherence in shallow NV centers. Our framework could be extended to include electric noise via random Stark shifts by promoting  $\epsilon_{x,y,z}$(t) to stochastic time-dependent fields that induce random Stark shifts at the NV. Other complementary approaches that explicitly model charge fluctuations and their impact on NV dephasing and relaxation can be found in Ref \cite{candido2024theory}. Additionally, we did not model explicit spin–strain or spin–stress couplings, instead, symmetry breaking is captured via the intrinsic transverse ZFS E splitting. In future work, one could compute surface and depth-dependent strain tensors from first principles for specific surface terminations and incorporate them into the NV spin-strain/stress Hamiltonian to quantify their contributions to level splittings and decoherence.

\section{Differences in T$_2$ values due to surface orientation}
\begin{figure*}[htbp!]
    \centering
    \includegraphics[width=0.9\textwidth]{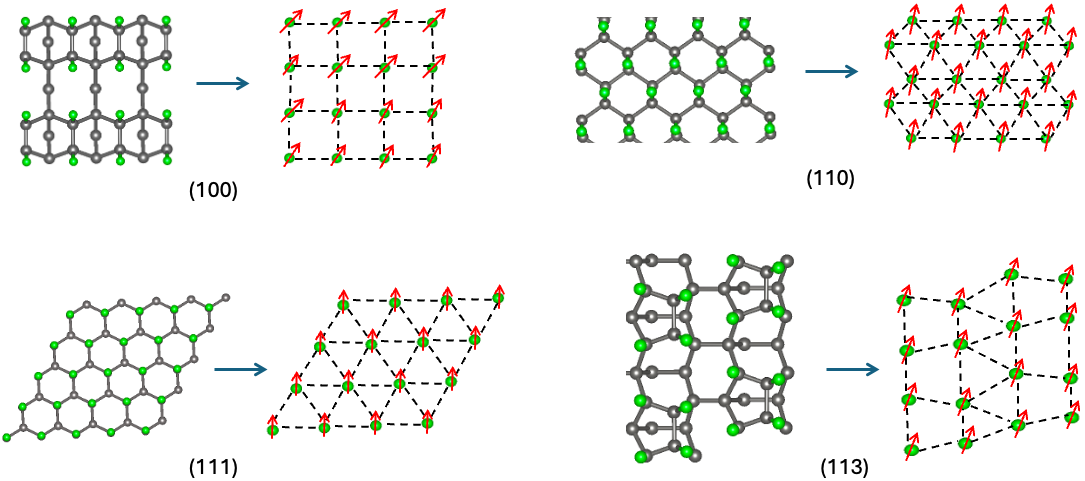}
    \caption{Flourine nuclear spin lattices for different fluorine-terminated surfaces: The crystallographic orientation of the diamond surface determines the geometry of the surface spin lattice: the (100) surface forms a square lattice, the (111) surface forms a triangular lattice, the (110) surface yields a rectangular (brick-wall) lattice, and the (113) surface gives rise to an oblique trapezoidal lattice. }
    \label{fig:Figure10}
\end{figure*}

\begin{figure*}[htbp!]
    \centering
    \includegraphics[width=0.9\textwidth]{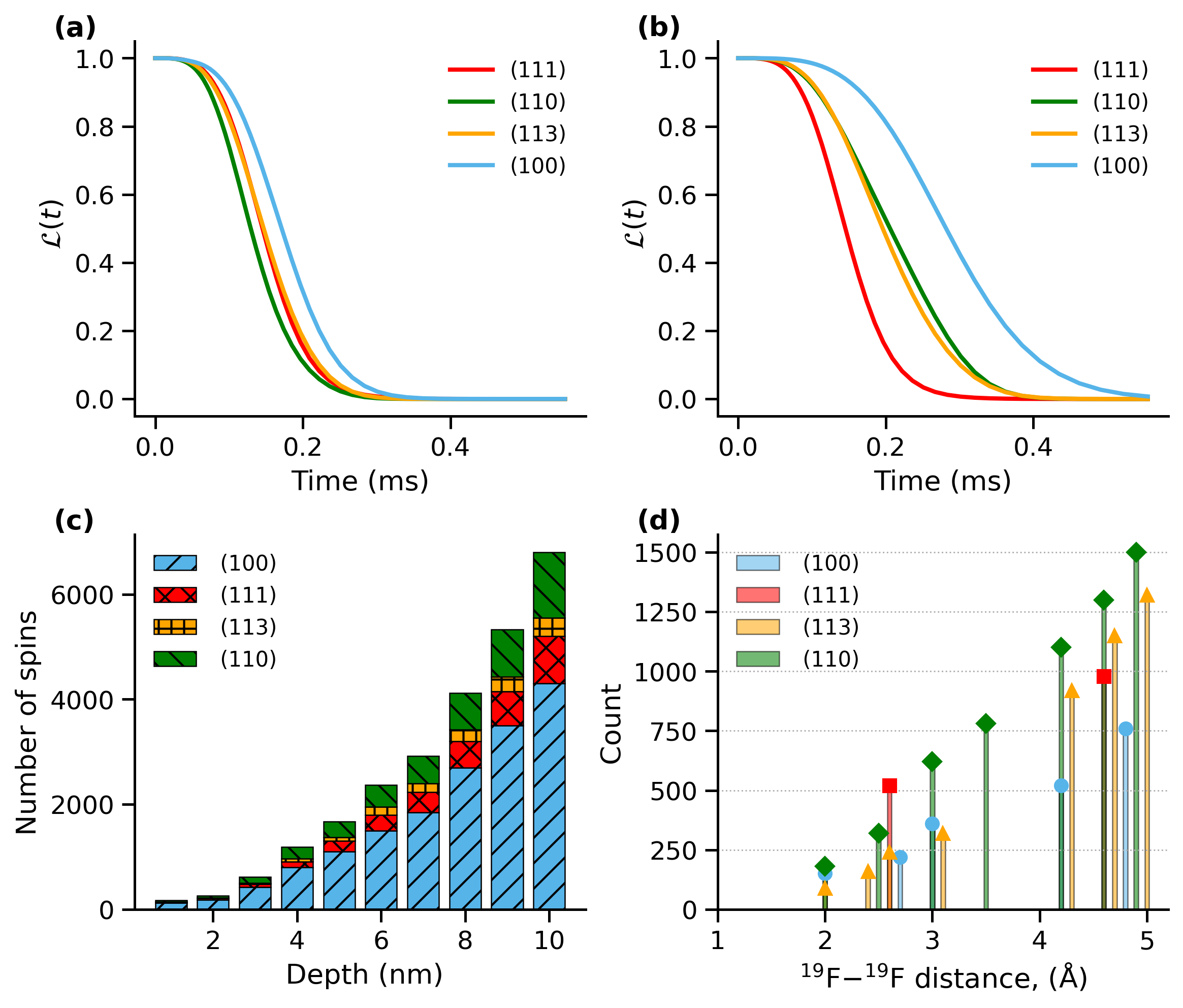}
    \caption{\textbf{(a)} Simulated spin-echo coherence $\mathcal{L}(t)$ for a near-surface NV$^-$ at depth $d_{\rm NV}=5$ nm, with the NV quantization axis parallel to the direction perpendicular to the surface. With the axis identical relative to its local surface in every case, the ordering of decays isolates the effect of surface-spin arrangement and areal density.\textbf{(b)} Coherence function $\mathcal{L}(t)$ at depth $d_{\rm NV}=5$ nm with the NV quantization axis fixed along the $\langle111\rangle$ direction for all surface orientation; axis subtends $54.7^\circ$ for the (100), $35.3^\circ$ for (110), $0^\circ$ for (111), and $29.5^\circ$ for (113) surfaces. The resulting decays reflect both lattice geometry/density and the orientation-dependent dipolar anisotropy.\textbf{(c)} Number of surface $^{19}$F spins within a fixed near-surface interaction volume as a function of NV depth for several surface orientations, showing orientation-dependent areal densities.
\textbf{(d)} Distribution of $^{19}$F–$^{19}$F separations across surface orientation.}
    \label{fig:Figure11}
\end{figure*}


For a fully fluorine-terminated diamond surface, $^{19}$F (nuclear spin $I = 1/2$) is the sole stable isotope and occurs with 100\% natural abundance. Consequently, complete surface termination corresponds to a monolayer of fluorine nuclear spins. As discussed in the main text, distinct surface orientations yield different $T_2$ values within the first 1–8 nm. These variations arise from the orientation-dependent arrangement of fluorine nuclei on the surface, which leads to differences in fluorine–fluorine internuclear spacing as well as in two-dimensional density of spins (spins per nm$^{-2}$). A comprehensive understanding of these effects requires consideration of both (i) the orientation of the NV center quantization axis relative to the surface normal, and (ii) the variation in spin arrangement and density across different crystallographic terminations. 

Fig.\ \ref{fig:Figure10} illustrates how the crystallographic orientation dictates the lattice geometry of fluorine nuclear spins: the (100) surface produces a rectangular (square-like) lattice, the (111) surface yields a triangular lattice, the (110) surface yields a rectangular/brick-wall lattice, while the (113) surface results in a trapezoidal arrangement. These orientation-dependent nuclear spin configurations represent realizations of distinct two-dimensional spin lattices and are relevant for the understanding of the NV decoherence. Indeed, previous theoretical work has shown that fluorine-terminated diamond surfaces can act as genuine quantum simulators of strongly correlated spin systems \cite{cai2013large}. Here we provide further evidence that different lattice geometries exhibit markedly different sensitivities to the NV $T_2$ coherence times within the first 1–8 nm. This result has a two-fold significance: on the one hand, the surface spin arrangement dictates the density and dipolar interaction strength of the fluorine nuclei, thereby influencing the NV decoherence; on the other hand, the spin arrangement networks constitute tunable platforms for simulating frustrated magnetism and related quantum phases at room temperature. Thus, our findings reinforce the proposal that fluorine-terminated diamond surfaces serve as experimentally accessible quantum simulators \cite{cai2013large}, while simultaneously revealing how NV coherence encodes the microscopic geometry of the underlying spin lattice.

\section{Electron spins on diamond Surface}
\subsection{Variance in coherence time calculations}
We find that the computed coherence of the electron spins do possess a large variance for dilute electron spin densities on the surface,  and when the NV centers are close to the surface. In general, the variance decreases as either the surface spin density increases (self-averaging over many bath spins) or as the NV depth increases (weaker coupling to the surface bath). 
For example, at a surface spin density of $\rho=0.001 ~nm^{-2}$, the standard deviation of the calculated $T_2$ reduces from $230 \ \mu s$ at 1 nm to $140 \ \mu s$ at 50 nm. 
At large depths, the residual variance is dominated by the stochastic configuration of 
$^{13} C$ nuclear spins in the diamond bulk lattice, consistent with prior CCE studies in bulk diamond that reported comparable spread across $\sim$ 1000 random 
 $^{13} C$ realizations \cite{maze2008electron}. Conversely, for high surface spin densities, we observe the opposite depth trend: as the NV depth increases, the influence of the surface bath diminishes and the relative contribution of the  $^{13} C$  bath becomes more prominent, leading to an increase in the $T_2$ variance with depth. Such configuration-induced variance and its suppression via self-averaging at higher spin densities have also been observed experimentally and reproduced in cluster-spin dynamical mean-field theory (CspinDMFT) simulations of interacting surface spin ensembles~\cite{GraesserPRR2023}. These results highlight the need for careful configurational averaging (over both surface spins and $^{13} C$ distributions) when computing NV coherence times near surfaces.
 
\begin{figure}[H]
    \centering
    \includegraphics[width=0.45\textwidth]{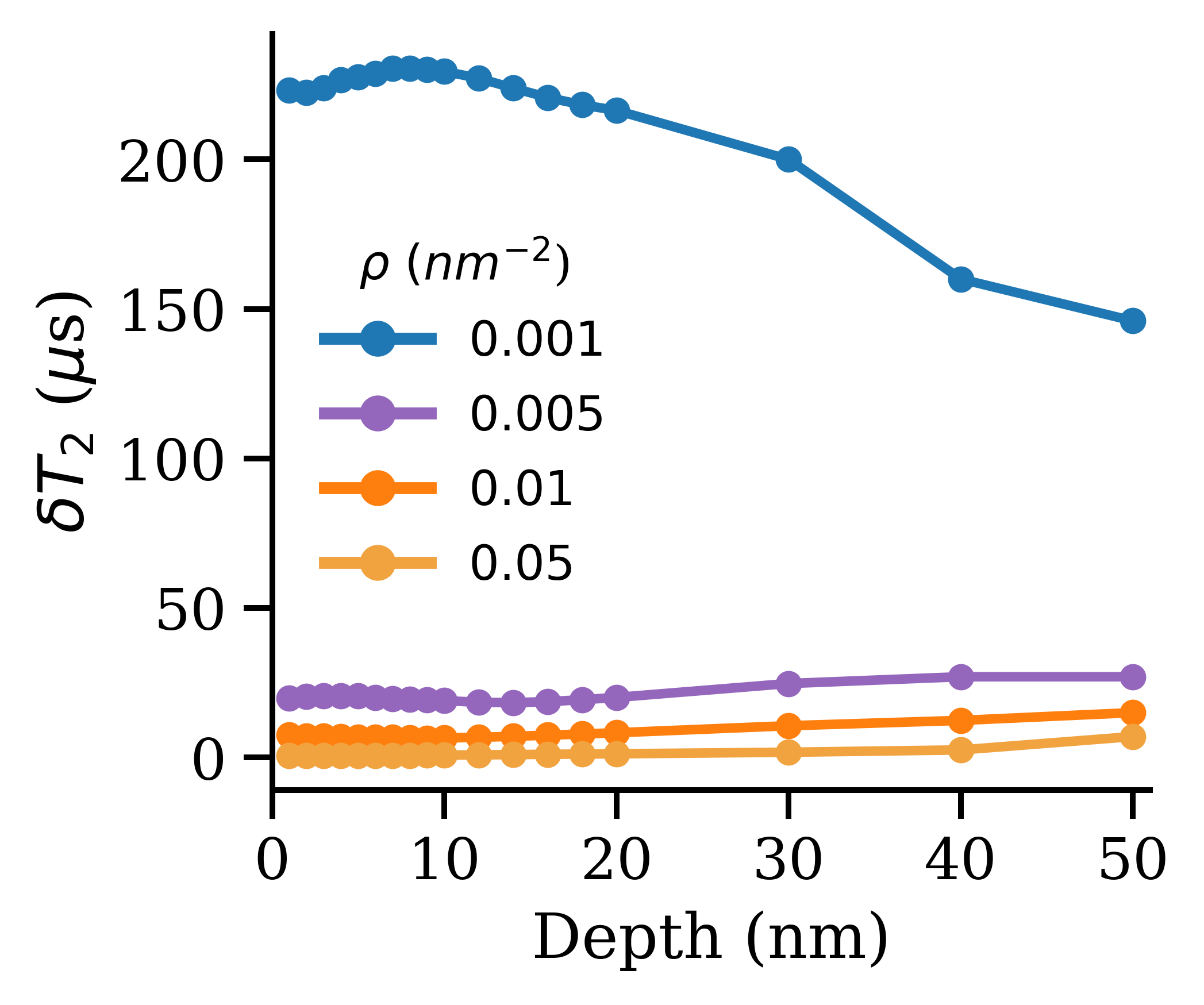}
    \caption{Standard deviation of the NV coherence times $\delta T_2$ as the function of NV centers' depth and surface electron spin density $\rho$.}
    \label{fig:Figure14}
\end{figure}

\subsection{Temperature Quenching of dipolar interaction in the static electron picture}
We further studied how temperature suppresses the strong dipolar interactions of localized surface spins, which, because of their large gyromagnetic ratios, dominate NV center decoherence. We addressed the following question: for a given surface spin density and NV depth, at what temperatures can one thermally quench flip-flop mediated dipolar noise from the surface spins? Although pulse‐based sequences such as WAHUHA \cite{WAHUHA} can also decouple dipolar couplings by Hamiltonian averaging, here we focus on thermal quenching via spin polarization.

To include the effect of temperature in our spin dynamics, we initialize each surface electron spin in a Gibbs state at a given temperature $T$:

\begin{equation}
   \rho = \frac{1}{Z}\sum_{m_s=-\frac{1}{2}}^{+\frac{1}{2}} e^{-\beta E_{m_s}}\,\ket{m_s}\bra{m_s}
\end{equation}

where $E_{m_s}=\gamma_sB_z(m_s)$, is the Zeeman energy of the spin projection $m_s$ on the field $B_z$,   $\beta = \frac{1}{k_B T}$  is the inverse thermal energy and  $Z=\sum_{m_s=-\frac{1}{2}}^{+\frac{1}{2}} e^{-\beta \gamma_s^NB_z(m_s)}$ is the partition function that normalizes $\rho$. 

\begin{figure}[H]
    \centering \includegraphics[width=0.5\textwidth]{ 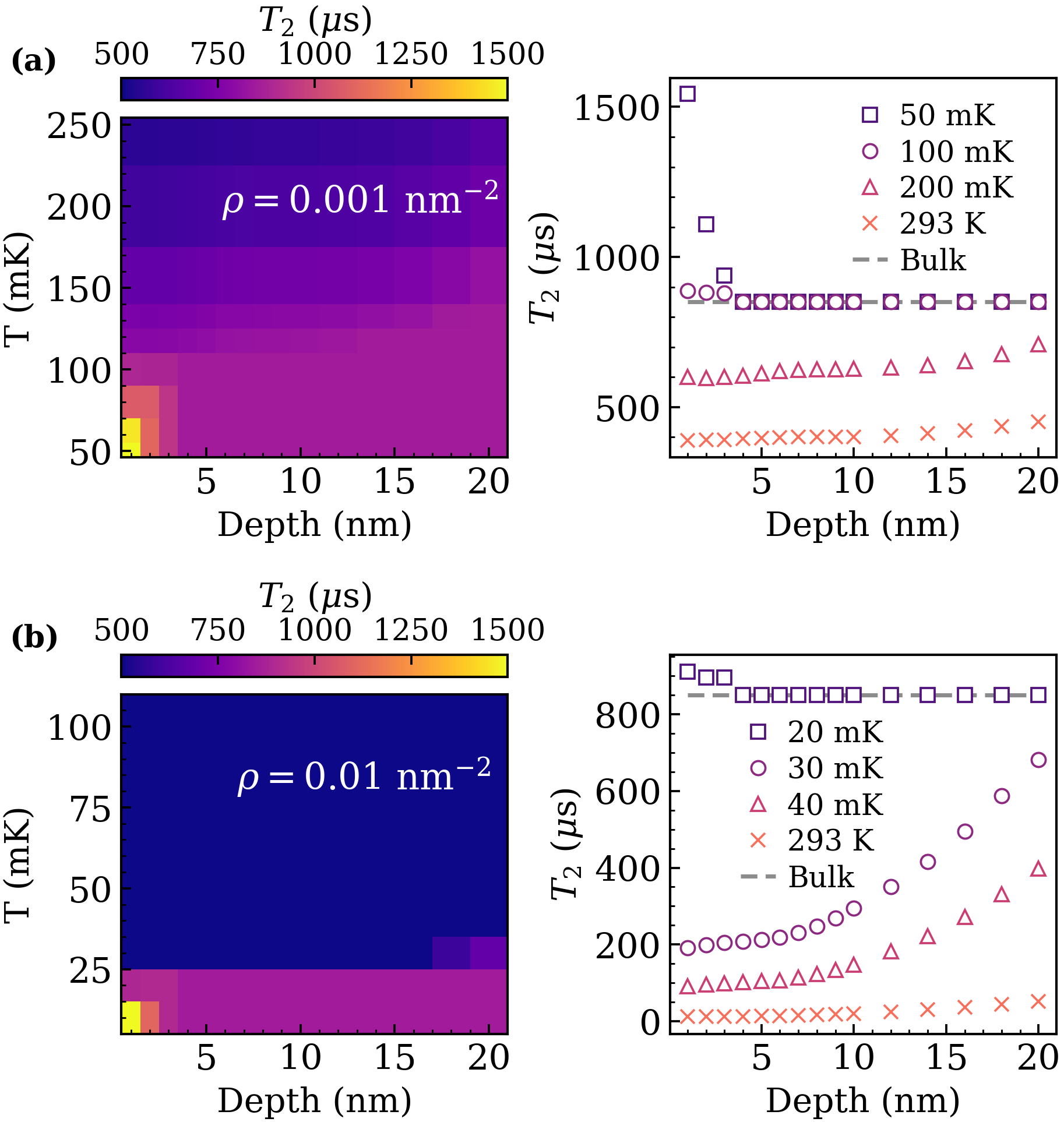}
   \caption{Contour maps (left panels) of computed Hahn-echo $T_2$ values as a function of temperature and NV centers' depths, for two surface spin densities: 0.001 nm$^{-2}$ \textbf{(a)} and 0.01 nm$^{-2}$ \textbf{(b)}. The corresponding $T_2$ as a function of depths and for bulk diamond are reported on the right panels.}
    \label{fig:temp}
\end{figure} 

As shown in Fig.~\ref{fig:temp}(a), for a surface density of \(\rho=0.001\ \mathrm{nm}^{-2}\), cooling the system to \(T=50\,\mathrm{mK}\) polarizes the surface spins and quenches flip–flop mediated dipolar noise at all depths. At \(T=100\,\mathrm{mK}\), this quenching is complete for \(d\ge4\)\,nm. The slight enhancement of \(T_2\) above the bulk value at \(d=1\text{–}3\)\,nm (for 50 and 100\,mK) reflects the lower nuclear‐spin density near the surface (see Fig.~\ref{fig:FIG2}). At larger depths, \(T_2\) saturates at the bulk-limited value (\(\sim850\,\mu\mathrm{s}\)), indicating that any residual decoherence is dominated by the presence of \(^{13}\)C nuclear spins.  
 For a higher density \(\rho=0.01\ \mathrm{nm}^{-2}\), as shown in Fig.~\ref{fig:temp}(b), the effective quenching occurs only at \(T=20\,\mathrm{mK}\) where flip–flop–mediated dipolar noise is suppressed. Hence we have identified the temperatures required to  suppress surface electronic-spin noise. At these temperatures: (i) NV centers positioned within a few nanometers of the diamond surface can achieve bulk‐like coherence times, and (ii) detection and characterization of proximal nuclear spins is achievable using nanoscale NMR. This result highlight pathways to optimize NV coherence through cryogenic operation.

\section{Motional narrowing of surface spins with finite relaxation}
\begin{figure*}[htp!]
    \centering
    \includegraphics[width=\textwidth]{ 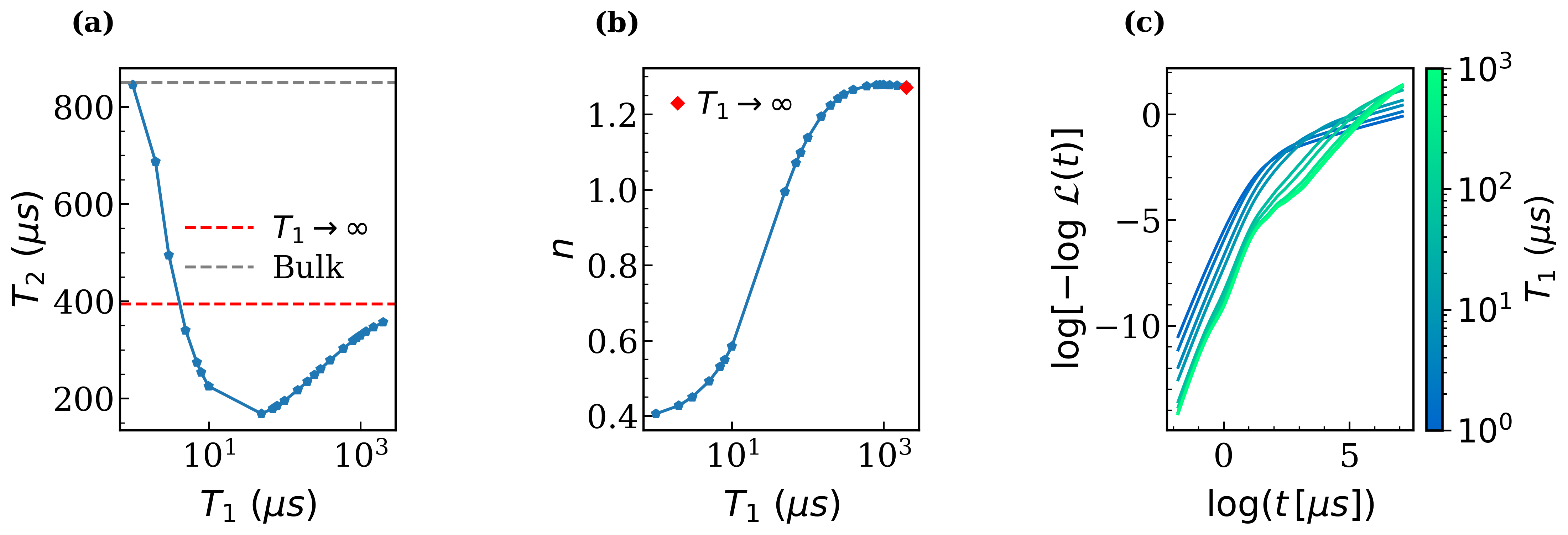}
   \caption{ \textbf{(a)} Computed Hahn–echo coherence time $T_2$ of NV center at a 5 nm distance from the surface, as a function of the surface‑spin relaxation time $T_1$, showing bulk-limited coherence (due to $^{13}$C) at $T_1= 1 \ \mu s$, a pronounced minimum near $T_1 = 50 \ \mu s$, and recovery toward the infinite-$T_1$ limit at longer relaxation times \textbf{(b)}. Stretch exponent $n$ extracted from the Hahn-echo decay as a function of $T_1$ \textbf{(c)}. Instantaneous stretch exponent obtained from the log-log coherence decays, demonstrating the progressive straightening of the decay curves at slow spin-flip rates. }
    \label{fig:FIG8}
\end{figure*}

To identify the regime where motional narrowing can suppress surface-induced decoherence and restore bulk $^{13}C$ limited coherence, we simulated a $5 \ nm$ deep NV center while varying the surface-spin relaxation time $T_1$ from $1$ to $ 1000 \ \mu s$ (see Fig.~\ref{fig:FIG8}(a)).

We find that at $T_1 = 1~\mu s$, the NV coherence is completely bulk‑limited, showing no influence from surface spins. As $T_1$ increases beyond $5 \ \mu s$, the decoherence induced by surface spin due to finite relaxation leads  to a decrease of  $T_2$ relative to its value in the infinite-$T_1$ limit, reaching a minimum near $T_1 \approx50 \ \mu s$.  Further increases in $T_1$ allow for $T_2$ to approach its infinite-relaxation limit. The stretched exponent (Fig.~\ref{fig:FIG8}b) also increases rapidly, approaching its infinite-$T_1$ value, while the instantaneous stretched exponent becomes more uniform, straightening the decay curve on a log-log scale (Fig.~\ref{fig:FIG8}c).

\section{Hopping dynamics}
To incorporate hopping dynamics within our spin bath, we map each bath spin as an effective spin-1 (three-level) system, where $\ket{+1}$ and $\ket{-1}$ represent the physical spin-1/2 states, while $\ket{0}$ encodes the hole (vacancy).  Importantly, the \emph{bath
spin operators are the same as those of spin-$\tfrac12$}: we consider 
$\mathbf{s}_i \equiv \boldsymbol{\sigma}_i/2$, which act non trivially only in the
$\{\ket{+1},\ket{-1}\}$ subspace and annihilate the hole state
($\mathbf{s}_i\ket{0_i}=0$). Thus we are not modeling an explicit spin-1 algebra for the bath.


In the ME-CCE framework \cite{onizhuk2024understanding}, incoherent hopping is included via pairwise Lindblad jump operators that
move a spin $s\in\{+1,-1\}$ from site $i$ into a hole at $j$:

\begin{equation}
  L_{i\to j}^{(s)} \equiv \ket{0_i\, s_j}\!\bra{s_i\, 0_j},
\end{equation}
affects the transformation $\ket{s_i\,0_j}\to\ket{0_i\,s_j}$.
For completeness, we also define local create/annihilate maps
\begin{equation}
  K_{+}^{(i,s)} \equiv \ket{s_i}\!\bra{0_i}, \qquad
  K_{-}^{(i,s)} \equiv \ket{0_i}\!\bra{s_i},
\end{equation}
so that $L_{i\to j}^{(s)} = K_{-}^{(i,s)}\otimes K_{+}^{(j,s)}$;  $K_{+}^{(i,s)}$ creates a spin-$s$ from a hole at site $i$, while
$K_{-}^{(i,s)}$ removes a spin-$s$ to a hole.

We consider spin-independent, symmetric (no directional bias) hopping rates $\Gamma_{i\to j}^{(s)}$, so that the forward and
backward rates are equal and independent from $s$:
\begin{equation}
  \Gamma_{i\to j}^{(s)} \;=\; \Gamma_{j\to i}^{(s)} \;\equiv\; \Gamma_{ij}.
\end{equation}
The pair rate decays with inter-site separation $r_{ij}$ between the spin at site $i$ and hole at site $j$ as
\begin{equation}\label{eq:lindblad_rate}
  \Gamma_{ij} \;=\; \frac{1}{t_{\mathrm{hop}}}\,
  \exp\!\left(-\frac{r_{ij}}{r_{\mathrm{hop}}}\right),
  \qquad r_{ij} \equiv \lvert \mathbf{r}_i - \mathbf{r}_j \rvert .
\end{equation}

\begin{figure}[H]
    \centering
    \includegraphics[width=0.42\textwidth]{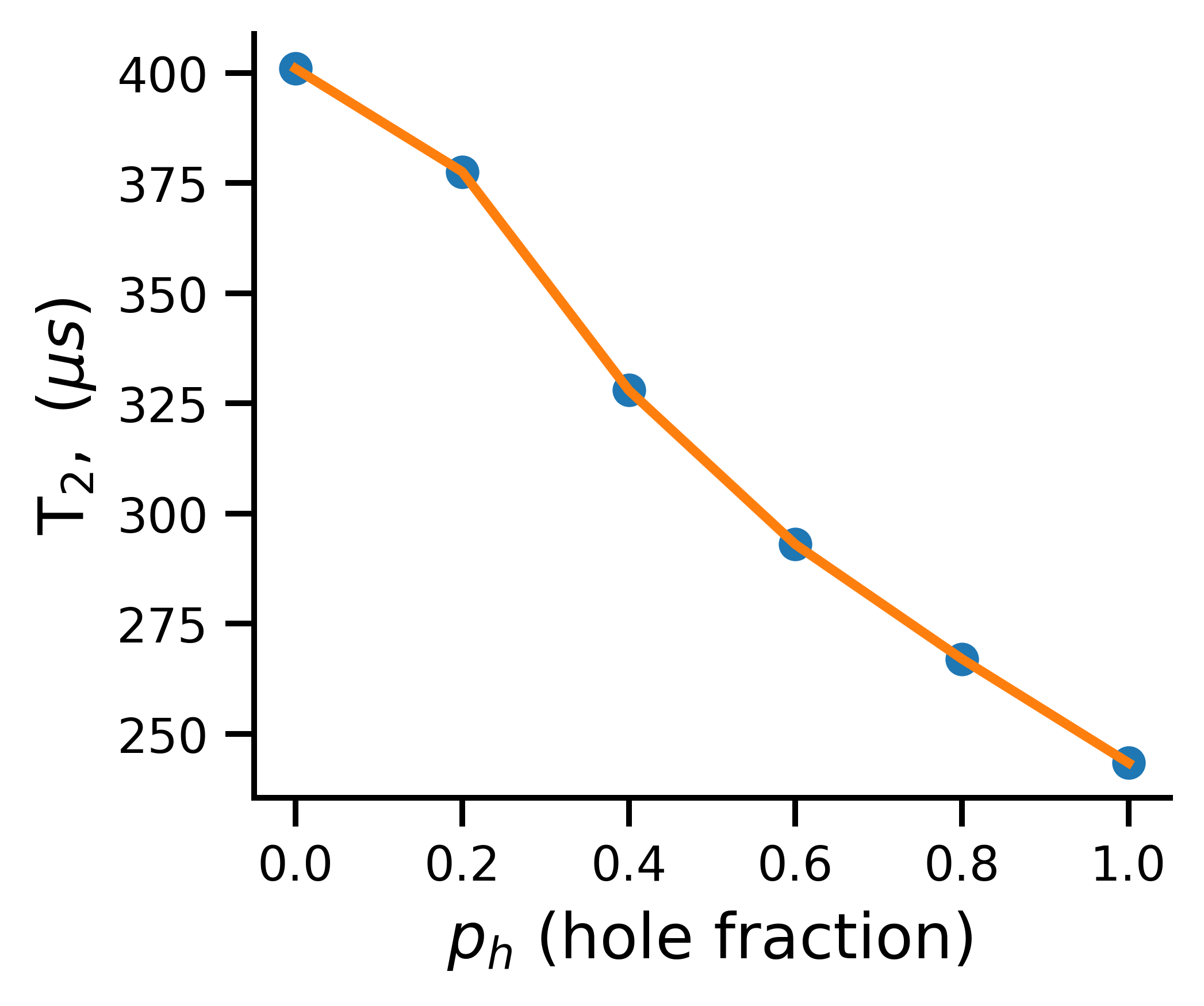}
    \caption{Spin-echo coherence time 
$T_2$ as a function of hole fraction $p_h$. Here 
$p_h \in [0,1]$ is the fraction of lattice sites designated as holes ($\rho^{holes}_i$).}
    \label{fig:Figure15}
\end{figure}

\textbf{Initial state preparation}
Each bath spin is prepared as an \emph{incoherent} mixture within the
spin-$\{\ket{+1},\ket{-1}\}$ subspace, with no population in $\ket{0}$:
\begin{equation}
  \rho^{\mathrm{spin}}_i \;=\; \tfrac{1}{2}\big(\ket{+1_i}\!\bra{+1_i}
  + \ket{-1_i}\!\bra{-1_i}\big), \qquad
  \bra{0_i}\rho^{\mathrm{spin}}_i\ket{0_i}=0.
\end{equation}
Vacancies (\emph{holes}) are initialized in the pure state
\begin{equation}
  \rho^{\mathrm{hole}}_i \;=\; \ket{0_i}\!\bra{0_i}.
\end{equation}
For a configuration $\mathcal{C}$ that specifies which sites are spins or holes, the
initial bath state is the product
\begin{equation}
  \rho_{\mathrm{bath}}(0\,|\,\mathcal{C}) \;=\; \bigotimes_i \rho^{(\mathcal{C}_i)}_i,
\end{equation}
and we model a controllable hole density $p_h\in[0,1]$ by sampling (or averaging over)
configurations with a fraction $p_h$ of sites set to $\rho^{\mathrm{hole}}_i$.
Since hopping occurs only into $\ket{0}$ levels, the hole density $p_h$ directly controls the
number of available acceptor sites and thereby the total escape rate out of any given spin site. For independently placed holes at fixed spin concentration, the expected rate from site $i$ scales roughly as:  $\langle\Gamma_i \rangle = p_h \sum_{j\neq i}\Gamma_{ij}$. 
Hence the strength of the
incoherent hopping channel increases monotonically with $p_h$, and the associated spectral
diffusion in the bath increases and the NV spin-echo coherence $T_2$ decreases
with $p_h$ (Fig.~\ref{fig:Figure15}). Consistent with this picture, prior experimental
work inferred that a large fraction of bath spins are mobile on the experimental
timescale, with an estimate of $\sim\!80\%$ participating in hopping processes~\cite{dwyer2022probing}.

\subsection{Experimental trends with static electron spins}
\begin{figure}[H]
    \centering
    \includegraphics[width=0.5\textwidth]{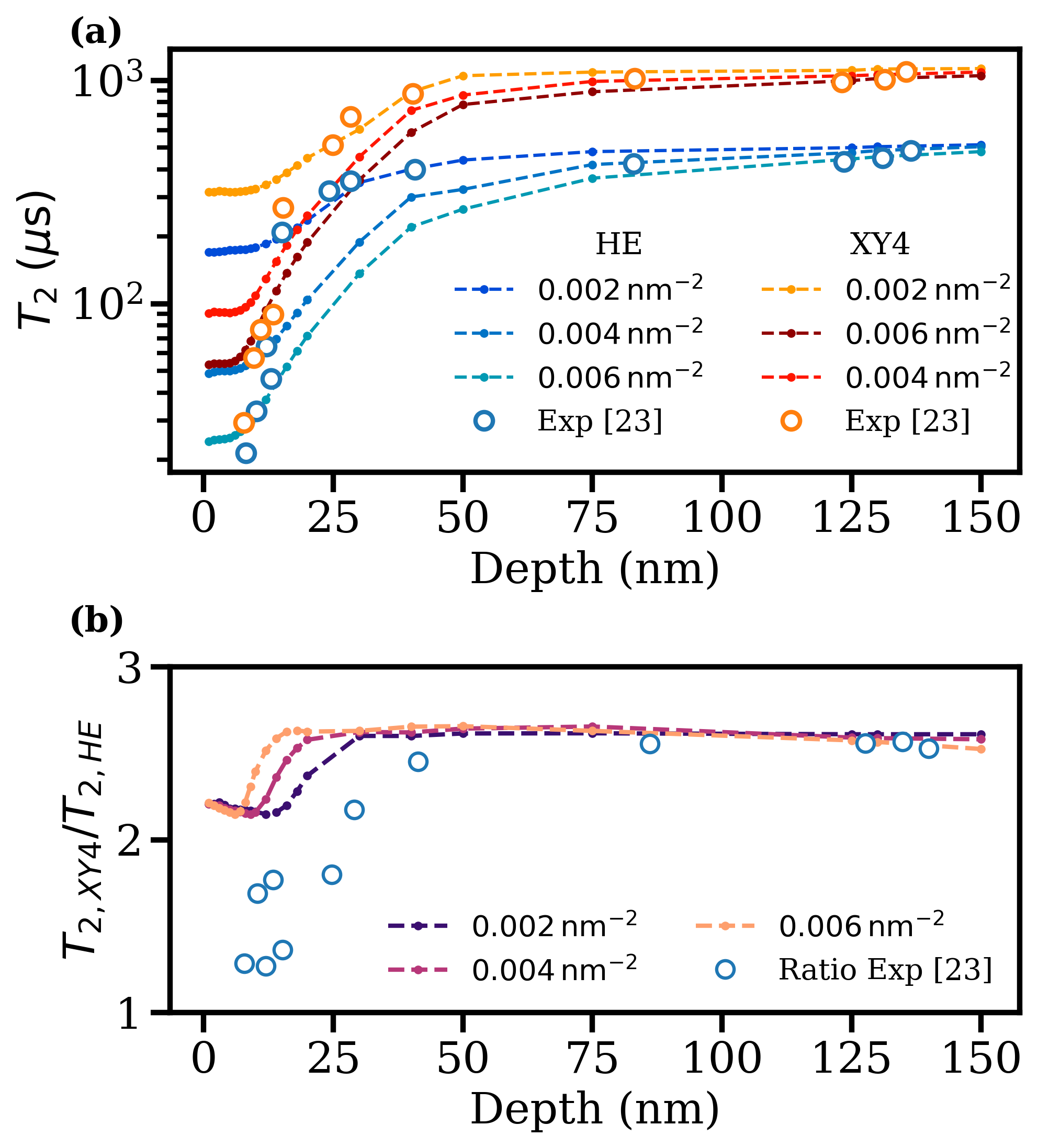}
    \caption{\textbf{(a)} Hahn-echo and XY-4 coherence time $T_2$  as a function of NV depth. Static surface-spin models require different assumed spin densities to match different depth ranges and fail to reproduce the full experimental trend with a single density. \textbf{(b)} Ratio of XY-4 to Hahn-echo coherence times as a function of NV centers’ depth. We only recover the experimental trend beyond 50 nm  }
    \label{fig:Figure16}
\end{figure}

Without hopping dynamics, the measured Hahn-echo $T_{2, HE}$ and the dynamical decoupled XY-4 $T_{2, XY4}$  as a function of NV depth behavior of near-surface NV centers cannot be reproduced with a single, static surface-spin density. As shown in Fig.~\ref{fig:Figure16}, matching the short-depth ($\lesssim 25$ nm) trends requires a much higher surface electron–spin density than is needed at intermediate depths, and only beyond $\sim 50$ nm do all curves approach the same bulk $T_2$ limit. This further indicate that dynamical changes in the surface spin bath or occasional hopping of surface spins is required to capture the experimentally observed behavior.

\nocite{*}
\clearpage

\bibliography{apssamp}

\end{document}